\documentclass[twocolumn,superscriptaddress,prl,amsmath,amstex,amssymb,citeautoscript,longbibliography]{revtex4-1}
\pdfoutput=1
\usepackage[subpreambles=true]{standalone}
\usepackage{rotating}
\usepackage{natbib}
\usepackage[english]{babel}
\usepackage{letltxmacro}
\usepackage{latexsym}
\usepackage[utf8]{inputenc}
\LetLtxMacro{\ORIGselectlanguage}{\selectlanguage}
\makeatletter
\DeclareRobustCommand{\selectlanguage}[1]{%
  \@ifundefined{alias@\string#1}
    {\ORIGselectlanguage{#1}}
    {\begingroup\edef\x{\endgroup
       \noexpand\ORIGselectlanguage{\@nameuse{alias@#1}}}\x}%
}
\newcommand{\definelanguagealias}[2]{%
  \@namedef{alias@#1}{#2}%
}
\makeatother
\definelanguagealias{en}{english}
\definelanguagealias{English}{english}
\usepackage{graphicx}
\usepackage{amsmath}
\usepackage{amsfonts}
\usepackage{amsthm}   
\usepackage{amssymb}
\usepackage{bm}
\usepackage{color}
\usepackage[percent]{overpic}
\usepackage{soul} 
\usepackage{amssymb}
\usepackage{wasysym}
\usepackage{dsfont}
\usepackage{float}

\usepackage{hyperref}
\hypersetup{
    bookmarks=false,         
    unicode=false,          
    pdftoolbar=false,        
    pdfmenubar=true,        
    pdffitwindow=false,     
    pdfstartview={FitH},    
    pdftitle={},    
    pdfauthor={Authors},     
    pdfsubject={},   
    pdfcreator={},   
    pdfproducer={}, 
    pdfkeywords={quantum many-body scars} {eigenstate thermalization hypothesis} {non-equilibrium dynamics}, 
    pdfnewwindow=true,      
    colorlinks=true,       
    linkcolor=black,          
    citecolor=blue,        
    filecolor=magenta,      
    urlcolor=blue           
}

\newcommand{\be}{\begin{equation}}
\newcommand{\ee}{\end{equation}}
\newcommand{\bea}{\begin{eqnarray}}
\newcommand{\eea}{\end{eqnarray}}

\usepackage{graphicx}
\usepackage[colorinlistoftodos]{todonotes}
\usepackage{verbatim}

\newcommand{\abs}[1]{\lvert#1\rvert}

\usepackage{soul}
\usepackage[normalem]{ulem}

\newtheorem*{lemma*}{Lemma}

\begin{abstract}
Recent experiments on Rydberg atom arrays have found evidence of anomalously slow thermalization and persistent density oscillations, which have been interpreted as a many-body analog of the phenomenon of quantum scars.
Periodic dynamics and atypical scarred eigenstates originate from a ``hard'' kinetic constraint: the neighboring Rydberg atoms cannot be simultaneously excited. Here we propose a realization of quantum many-body scars in a 1D bosonic lattice model with a ``soft'' constraint in the form of density-assisted hopping. We discuss the relation of this model to the standard Bose-Hubbard model and possible experimental realizations using ultracold atoms. We find that this model exhibits similar phenomenology to the Rydberg atom chain, including weakly entangled eigenstates at high energy densities and the presence of a large number of exact zero energy states, with distinct algebraic structure.

\end{abstract}

\begin{document}

\title{Quantum scars of bosons with correlated hopping}
 \author{Ana Hudomal}
\email[]{ana.hudomal@ipb.ac.rs}
\affiliation{Institute of Physics Belgrade, University of Belgrade, 11080 Belgrade, Serbia}
\author{Ivana Vasi\'c}
\affiliation{Institute of Physics Belgrade, University of Belgrade, 11080 Belgrade, Serbia}
\author{Nicolas Regnault}
\affiliation{Joseph Henry Laboratories and Department of Physics, Princeton University, Princeton, New Jersey 08544, USA}
\affiliation{Laboratoire de Physique de l'Ecole Normale Sup\'{e}rieure, ENS, Universit\'{e} PSL, CNRS, Sorbonne Universit\'{e}, Universit\'{e} Paris-Diderot, Sorbonne Paris Cit\'{e}, 75005 Paris, France}
\author{Zlatko Papi\'c}
\affiliation{School of Physics and Astronomy, University of Leeds, Leeds LS2 9JT, United Kingdom}
\date{\today}
\maketitle


Semiclassical studies of chaotic stadium billiards have revealed the existence of remarkable non-chaotic eigenfuctions called ``quantum scars''~\cite{Heller84}. Scarred eigenfunctions display anomalous enhancement in regions of the billiard that are traversed by one of the periodic orbits in the classical limit when $\hbar\to 0$. It was shown that quantum scars lead to striking experimental signatures in a variety of systems, including microwave cavities~\cite{Sridhar1991}, quantum dots~\cite{Marcus1992}, and semiconductor quantum wells~\cite{Wilkinson1996}. 

A recent experiment on a quantum simulator~\cite{Bernien2017}, and subsequent theoretical work~\cite{Turner2017,wenwei18TDVPscar}, have shown that quantum many-body scars can occur in strongly interacting quantum systems. The experiment used a one-dimensional Rydberg atom platform in the regime of the Rydberg blockade~\cite{Schauss2012,Labuhn2016, Bernien2017}, where nearest-neighbour excitations of the atoms were energetically prohibited. The experiment observed persistent many-body revivals of local observables after a ``global quench"~\cite{CalabreseQuench} from a certain initial state. In contrast,  when the experiment was repeated for other initial configurations, drawn from the same type of ``infinite" temperature ensemble, the system displayed fast equilibration and no revivals. These observations pointed to a different kind of out-of-equilibrium behavior compared to previous studies of quantum thermalization in various experimental platforms~\cite{Kinoshita06, Bloch15, Monroe16, Lukin16, Kaufman2016}. 

In both single-particle and many-body quantum scars, the dynamics from certain initial states leads to periodic revivals of the wave function. In the former case, this happens when the particle is prepared in a Gaussian wave packet initialized along a periodic orbit~\cite{Heller84}, while in the latter case the revivals can be interpreted as a nearly-free precession of a large emergent $\mathrm{su(2)}$ spin degree of freedom~\cite{Choi2018, Bull2020}.
Another similarity between single- and many-body quantum scars is the existence of non-ergodic eigenstates. In the single-particle case, such eigenstates are easily identified by their non-uniform probability density that sharply concentrates along classical periodic orbits. In the many-body case, non-ergodic eigenstates are broadly defined as those that violate Eigenstate Thermalization Hypothesis (ETH)~\cite{DeutschETH,SrednickiETH}. 
Scarred eigenstates violate the ETH in a number of ways: for example, they appear at evenly spaced energies throughout the spectrum~\cite{Turner2017,TurnerPRB,Iadecola2019}, they have anomalous expectation values of local observables compared to other eigenstates at the same energy density, and their entanglement entropy obeys a sub-volume law scaling~\cite{TurnerPRB}.  

In recent works, the existence of atypical eigenstates has been taken as a more general definition of quantum many-body scaring. For example, highly-excited eigenstates with low entanglement have previously been analytically constructed in the non-integrable AKLT model~\cite{Bernevig2017,BernevigEnt}. A few of such exact eigenstates are now also available for the Rydberg atom chain model~\cite{lin2018exact}. The collection of models that feature atypical eigenstates
is rapidly expanding, including perturbations of the Rydberg atom chain~\cite{Khemani2018, TurnerPRB, Michailidis2019}, theories with confinement~\cite{Calabrese16, Konik1, Konik2}, Fermi-Hubbard model beyond one dimension~\cite{Vafek, IadecolaZnidaric}, driven systems~\cite{Haldar2019}, quantum spin systems~\cite{Schecter2019,Iadecola2019b}, fractional quantum Hall effect in a one-dimensional limit~\cite{Moudgalya2019}, and models with fracton-like dynamics~\cite{Pretko2019,Khemani2019,Sala2019,Khemani2019_2}. In a related development, it was proposed that atypical eigenstates of one Hamiltonian can be ``embedded'' into the spectrum of another, thermalizing Hamiltonian~\cite{ShiraishiMori}, causing a violation of a ``strong'' version of the ETH~\cite{dAlessio2016, Gogolin2016}. This approach allows to engineer scarred eigenstates in models of topological phases in arbitrary dimensions~\cite{NeupertScars}. From a dynamical point of view, it has been shown that models with scarred dynamics can be systematically constructed by embedding periodic on-site unitary dynamics into a many-body system~\cite{Bull2019}.

A feature shared by many scarred models is the presence of some form of a kinetic constraint. In the Rydberg atom chain, the constraint results from strong van der Waals forces, which project out the neighboring Rydberg excitations~\cite{Lesanovsky2012}. Such Hilbert spaces occur, for example, in models describing anyon excitations in topological phases of matter~\cite{Feiguin07, Trebst2008, Chandran16,Lan2017, Chandran2019} and in lattice gauge theories~\cite{Lan2017_2,Smith2017,brenes2017many}, including the Rydberg atom system~\cite{Surace2019,Magnifico2019}. Recent works on periodically driven optical lattices have started to explore such physics~\cite{Gorg2018,Schweizer2019}. On the other hand, kinetic constraints have been investigated as a possible pathway to  many-body localization without disorder~\cite{AbaninRev}. In classical systems, non-thermalizing behavior without disorder is well-known in the context of structural glasses~\cite{Binder2011,Berthier2011,Biroli2013}. The mechanism of this type of behavior is the excluded volume interactions that impose kinetic constraints on the dynamics~\cite{Fredrickson1984,Palmer1984}.  Similar type of physics has recently been explored in quantum systems where a ``quasi many-body localized" behavior was proposed to occur in the absence of disorder~\cite{carleo2012localization, Huveneers13, Muller, Yao14, Papic, Juan15, Veness, Smith2017a, Kim2016, Yarloo2017, michailidis2017slow}. 

In this paper we investigate the relation between kinetic constraints, slow dynamics and quantum many-body scars. In contrast to previous work, which focused on models of spins and fermions that are closely related in one dimension due to the Jordan-Wigner mapping, here we study one-dimensional models of bosons with density-assisted hoppings, which realize both ``hard" and ``soft" kinetic constraints, whilst being non-integrable. Depending on the form of the hopping term, we demonstrate that the models encompass a rich phenomenology, including regimes of fast thermalization, the existence of periodic revivals and many-body scars, as well as the Hilbert space fragmentation that has been found in recent studies of fractonic models~\cite{Pretko2019,Khemani2019,Sala2019,Khemani2019_2}. Unlike the experimentally realized Rydberg atom system, we find evidence of many-body scars in a bosonic model without a hard kinetic constraint, i.e., with a fully connected Hilbert space. We identify initial states that give rise to periodic many-body revivals in the quantum dynamics, and we introduce a ``cluster approximation" that captures the scarred eigenstates that are responsible for periodic revivals. We discuss possible experimental realizations of these models using ultracold atoms.

 \begin{figure*}[htb]
\includegraphics[width=0.8\textwidth]{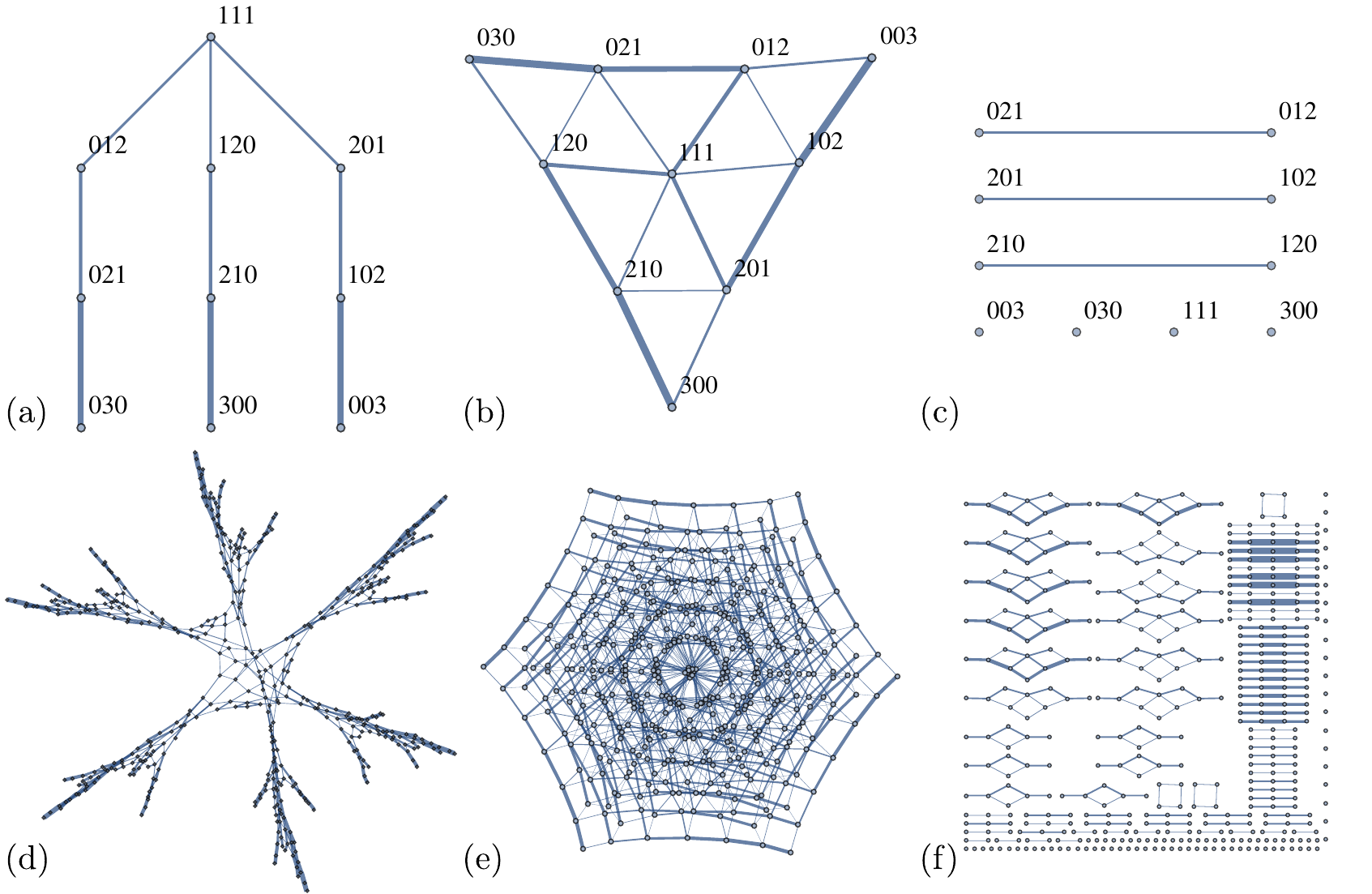}
\caption{Connectivity of the Hilbert space. 
Adjacency graph for (a) $H_1$, (b) $H_2$, (c) $H_3$, all for $L=N_\mathrm{p}=3$. (d), (e) and (f): same as (a), (b) and (c) but for  $L=N_\mathrm{p}=6$. To avoid clutter, we do not label the vertices in (d), (e) and (f).
All graphs are weighted, i.e., the line thickness is proportional to the magnitude of the corresponding hopping coefficient.
Several different clusters of configurations are visible in the case of $H_1$. The clusters start to form already for $L=3$ (for example, the configurations $012$-$021$-$003$ in (a)) and become more prominent for $L=6$ (d).
In the case of $H_2$, almost all configurations are well-connected to the rest of the graph. The graphs for $H_3$ show that the Hilbert space is highly reducible: its graph splits into many disconnected components. 
}\label{fig:graphs}
\end{figure*}

\section{Results}
\subsection{Models and their Hilbert spaces}\label{s:models}

A fundamental ingredient of kinetically constrained models is ``correlated hopping":  a particle can hop depending on the state of its neighbors. In this paper we consider a system of $N_\mathrm{p}$ bosons on a one-dimensional lattice with $L$ sites. We consider models where the total filling factor, $\nu=N_\mathrm{p}/L$, is conserved, and we will mainly present results in the dense regime, $\nu=1$. We have studied models with $\nu<1$ and $\nu>1$, but we found them to be either too constrained or not constrained enough, and therefore less interesting. We emphasize that the bosons in our study are not hard-core, i.e., the occupancy of any lattice site can take any value from $0$ to $N_\mathrm{p}$.

We study three different models, defined by the Hamiltonians: 
\begin{eqnarray}
\label{eq:h1}
 H_1&=&-J\sum_{j=1}^L \left( b^\dagger_{j}b_{j+1} n_{j}+n_{j-1}b^\dagger_{j} b_{j-1}\right),\\
 \label{eq:h2}
 H_2&=&-J\sum_{j=1}^L \left(n_{j}b^\dagger_{j}b_{j+1}+b^\dagger_{j}b_{j-1}n_{j-1}\right),\\
\label{eq:h3}
 H_3&=&-J\sum_{j=1}^L \left(n_{j+1}b^\dagger_{j}b_{j+1}n_{j}+n_{j-1}b^\dagger_{j}b_{j-1}n_{j}\right).
 \end{eqnarray}
All three models contain a free-boson hopping term, $b_j^\dagger b_{j+1}$, which is dressed in various ways by density operators, $n_{j}=b^\dagger_{j}b_{j}$.  We will show that the position of the density operator $n_j$ completely changes the behavior of these models, ranging from fast thermalization to the breakup of the Hamiltonian into disconnected, exactly solvable sectors. 
For example, note that $H_1$ and $H_2$ are related to each other via free boson hopping,
 \begin{eqnarray}\label{eq:h12}
 H_2&=& H_1-J\sum_{j} \left(b^\dagger_{j}b_{j+1}+b^\dagger_{j}b_{j-1}\right),
 \end{eqnarray}
 which can be easily proven using bosonic commutation relations. We will see below that this innocuous free-boson hopping leads to surprisingly different dynamical properties of the two models.

The motivation behind introducing three different models in Eqs.~(\ref{eq:h1})-(\ref{eq:h3}) can be summarized as follows. Hamiltonian $H_1$ describes a model where a particle cannot hop to the left if that site is not already occupied by at least one particle, and cannot hop to the right if it is the only particle left on its initial site. This introduces constraints to the system.
 Conversely, there are no such constraints in the case of $H_2$. Indeed, the hopping coefficients are only modified in intensity by the particle-number operator. 
 Hamiltonian $H_3$ introduces additional constraints compared to $H_1$. The number of unoccupied sites and their positions remain constant under the action of this Hamiltonian. This leads to different connectivity of the Hilbert space in each of the models, as we explain in the next Section.

We consider periodic boundary conditions ($L+1\equiv 1$) and set $\hbar=J=1$. With periodic boundary conditions, all three Hamiltonians $H_1$, $H_2$ and $H_3$ have translation symmetry, thus their eigenstates can be labelled by momentum quantum number, $k$, quantized in units of $2\pi/L$. In addition, $H_3$ has inversion symmetry. We denote by $I=0$ and $I=1$ the sectors that are even and odd under inversion, respectively. 

 Without restrictions on the boson occupancy, the Hilbert space of $H_1$, $H_2$ and $H_3$ grows very rapidly. For $L=N_\mathrm{p}=12$, the Hilbert space size of the $k=0$ sector is $112720$ (the largest one we will consider for $H_1$ and $H_2$). As previously mentioned (see also the next Section), 
 the Hilbert space of $H_3$ splits into many disconnected components, thus it is possible to consider only one connected component at a time and disregard the unoccupied sites whose positions do not change. 
 This is more relevant when looking at properties such as thermalization, than fixing the filling factor.
 However, the boundary conditions are in that case no longer periodic, and the system does not have translation symmetry. 
 Considering only a  system with the size $L/2$, filling factor $\nu=2$, open boundary conditions and minimal number of particles per site equal to $1$ is completely equivalent to considering the largest component of the full system which has the size $L$, filling factor $\nu=1$, periodic boundary conditions and no restrictions on the occupancies.
 The Hilbert space size of the symmetric invariant sector of 
 the largest connected component of $L=N_\mathrm{p}=22$ is $176484$ and this is the largest sector that we will consider for $H_3$.

\subsection{Graph structure of the models}\label{sec:graph}

Since we will be interested in the dynamical properties, it is convenient to first build some intuition about the structure of the Hamiltonians of the three models in Eqs.~(\ref{eq:h1})-(\ref{eq:h3}). A Hamiltonian can be viewed as the adjancency matrix of a graph whose vertices are Fock states of bosons, $|n_1,n_2,\ldots,n_L\rangle$. If the Hamiltonian induces a transition between two Fock states, the corresponding vertices of the graph are connected by a link. The graphs that show how the configuration space is connected 
have very different structure for the three Hamiltonians $H_1$, $H_2$ and $H_3$, as can be observed in Fig.~\ref{fig:graphs}. 

The entire graph of $H_2$ is well-connected and it has the same structure as the graph of the standard Bose-Hubbard model: the particle-number operators in $H_2$ do not introduce any constraints, but only affect the magnitude of the hopping coefficients. In contrast, the $H_1$ graph shows several clusters of configurations that are weakly connected to the rest of the graph. 
  ``Weakly connected'' means that there is a small number of connections leading outside the cluster and that their respective hopping coefficients are smaller in magnitude than those of the surrounding connections within the cluster. A state that is initially located inside a cluster is therefore more likely to stay inside during an initial stage of the time evolution, which increases the probability of revivals and slows down the growth of entanglement entropy. We will provide a more quantitative description and examples that illustrate this in Section ``Quantum scars in $H_1$ and $H_3$ models''. 
  Finally, the graph of $H_3$, due to even stronger constraints, is actually disconnected, which is an example of Hilbert space fragmentation that was previously shown to cause non-ergodic behavior in fracton-like models~\cite{Khemani2019,Sala2019}. This predicts that thermalization and dynamics in the three models will be very different, which we will confirm in the following Section. 
  However, we note that the number of connections and the topology of the graph is not the only relevant factor for the dynamics. The magnitude of the hopping coefficients between different configurations is also important  (Supplementary Note~1). 

 We note that the relation between $H_1$ and $H_3$ is reminiscent of the relation between the quantum East model~\cite{Garrahan15} and the ``PXP'' model describing the atoms in the Rydberg blockade regime~\cite{Lesanovsky2012, Turner2017, TurnerPRB}. Like $H_3$, the PXP model is doubly constrained and inversion symmetric, while $H_1$ and the quantum East model are asymmetric versions of those two models with only a single constraint. The graph of the quantum East model is similar to that of $H_1$, in that it contains bottlenecks which slow down the growth of entanglement entropy~\cite{Garrahan15}. 

\begin{figure*}[htb]
\includegraphics[width=0.99\textwidth]{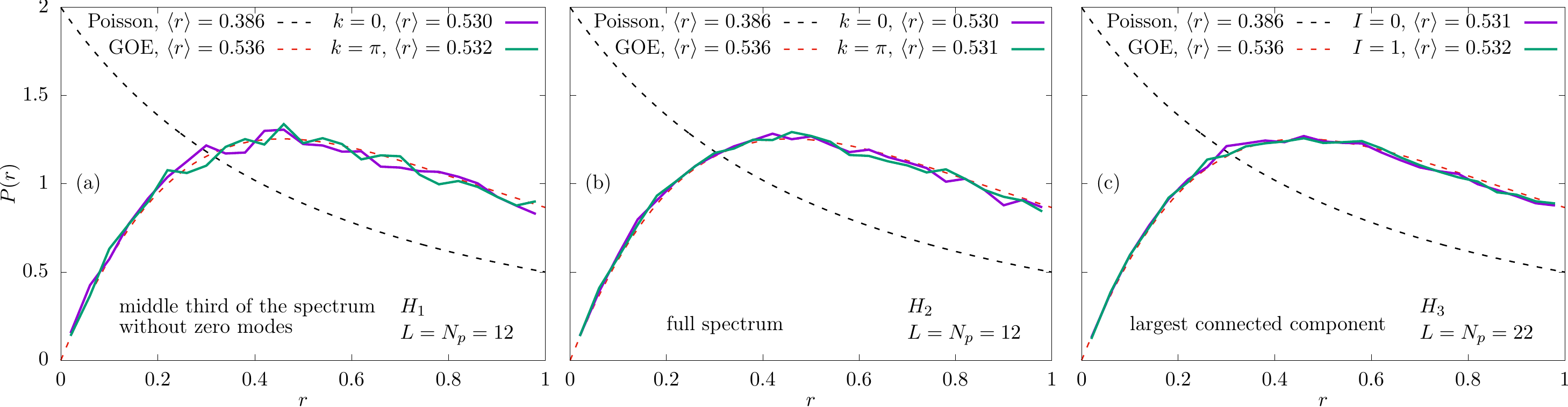}
\includegraphics[width=0.99\textwidth]{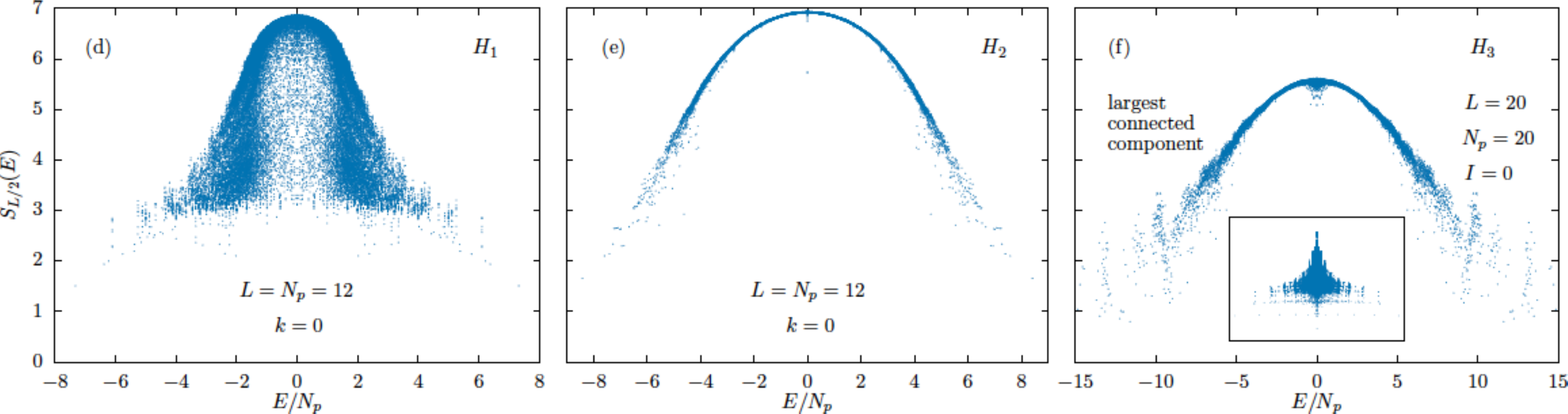}
\caption{ Level statistics and entanglement. (a), (b) and (c): Probability distribution of the ratios of two consecutive energy gaps. 
(a) $H_1$ (middle third of the spectrum without $E=0$ states, $L=N_\mathrm{p}=12$), (b) $H_2$ (full spectrum, $L=N_\mathrm{p}=12$) 
and (c) $H_3$ (largest connected component of $L=N_\mathrm{p}=22$).
The black dashed line shows the Poisson distribution, which corresponds to the integrable case, while the red dashed line is the distribution of the Gaussian orthogonal ensemble, which corresponds to the thermalizing case. 
(d), (e) and (f): Entanglement entropies $S_{L/2}$ of all eigenstates plotted as a function of the eigenstate energy per particle, $E/N_\mathrm{p}$. (d) $H_1$ ($L=N_\mathrm{p}=12$, $L_\mathrm{A}=6$, $k=0$), (e) $H_2$ (same) and (f) $H_3$ in the largest connected component of $L=N_\mathrm{p}=20$, $L_\mathrm{A}=10$, $I=0$.  
The inset shows all connected components for $L=N_\mathrm{p}=12$, $L_\mathrm{A}=6$, $k=0$.
}\label{fig:spacing}
\end{figure*}

\subsection{Dynamics and entanglement properties}\label{s:dynamics}

We now investigate the phenomenology of the models introduced in Eqs.~(\ref{eq:h1})-(\ref{eq:h3}). We use  exact diagonalization to obtain the complete set of energy eigenvalues and eigenvectors, from which we evaluate the level statistics and the distribution of entanglement entropies for the three models. Furthermore, we probe dynamical properties of the models by studying a global quench, simulated via Krylov iteration.  

The energy level statistics is a standard test for thermalization of models that cannot be solved exactly. A convenient way to probe the level statistics is to examine the probability distribution $P(r)$~\cite{OganesyanHuse} of ratios between consecutive energy gaps $s_n=E_{n+1}-E_n$, 
\begin{eqnarray}
r=\frac{\mathrm{min}(s_n,s_{n+1})}{\mathrm{max}(s_n,s_{n+1})}.
\end{eqnarray}
The advantage of studying $P(r)$, instead of $P(s_n)$, is that there is no need to perform the spectrum unfolding procedure -- see Ref.~\onlinecite{DAlessio2014}. For standard random matrix theory ensembles, both $P(r)$ and the mean $\langle r\rangle$ are well-known~\cite{Mehta2004}. When computing the same quantities in a microscopic physical model, it is crucial to resolve all the symmetries of the model.

The probability distribution $P(r)$ of the ratios  of two consecutive energy gaps  is shown in Figs.~\ref{fig:spacing}(a), (b) and (c) for the three Hamiltonians $H_1$, $H_2$ and $H_3$ respectively, and two 
momentum or inversion sectors. In all three cases, 
the energy levels repel, i.e., the distribution tends to zero as $r\to 0$. For $H_2$, the distribution is particularly close to the Wigner-Dyson (non-integrable) line. For $H_1$, the distribution is also consistent with Wigner-Dyson when we restrict to the middle 1/3 of the spectrum (and after removing special states with $E=0$). 
We exclude the edges of the spectrum because they contain degeneracies which are not symmetry-related. 
However, such states do not appear to have a major effect on the level statistics distribution, which is still closer to the Wigner-Dyson than the Poisson distribution even if they are included. 
The level statistics of $H_3$ within the largest connected component of the Hilbert space is shown in Fig.~\ref{fig:spacing}(c) and is also consistent with the Wigner-Dyson distribution without restricting the spectrum. 
However, we will demonstrate below that the dynamics in some smaller connected components of $H_3$ can be exactly solved.

 \begin{figure*}
\includegraphics[width=0.32\textwidth]{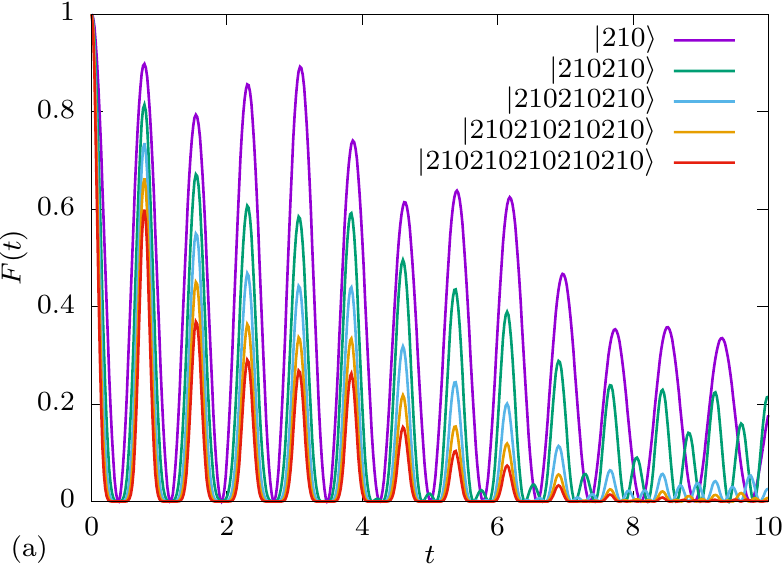}
\includegraphics[width=0.32\textwidth]{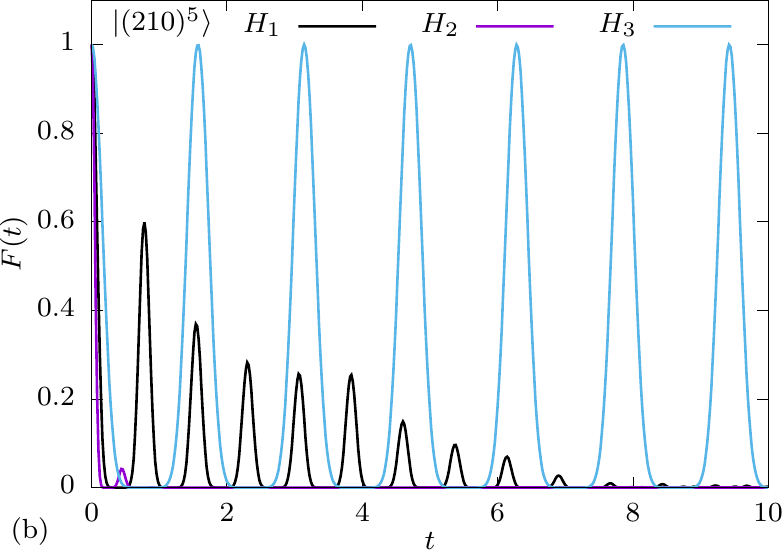}
\includegraphics[width=0.32\textwidth]{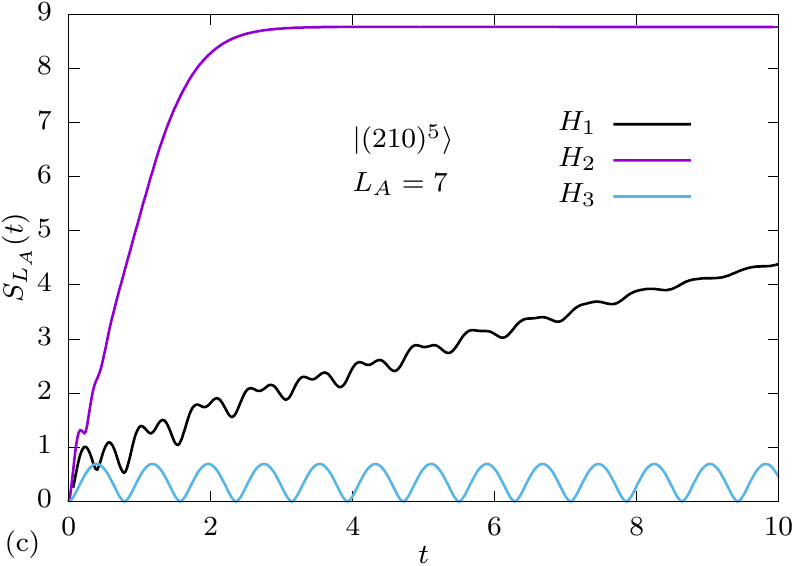}
\caption{Dynamics of quantum fidelity and entanglement entropy for initial configurations in Eq.~(\ref{eq:config}). 
(a) Time evolution of fidelity $F(t)$ in Eq.~(\ref{eq:fid}) for system sizes $L=3n$. The evolution is governed by the Hamiltonian $H_1$, different colors represent different system sizes $L$. 
(b) Fidelity evolution $F(t)$ for the Hamiltonians $H_1$, $H_2$ and $H_3$ and system size $L=15$.
(c) Entanglement entropy evolution $S_{L_\mathrm{A}}(t)$ for the same cases as in (b).
}\label{fig:dyn021}
\end{figure*}

As a complementary diagnostic of thermalization, we next compute the entanglement entropy of all eigenstates. We divide the lattice into two sublattices, A and B, of lengths $L_\mathrm{A}$ and $L_\mathrm{B}=L-L_\mathrm{A}$. For a given pure state $|\psi\rangle$, the entanglement entropy is defined as
\begin{eqnarray}\label{eq:ee}
S_\mathrm{A}=-\mathrm{tr}_\mathrm{A}(\rho_\mathrm{A}\ln\rho_\mathrm{A}),
\end{eqnarray}
where $\rho_\mathrm{A}=\mathrm{tr}_\mathrm{B}|\psi\rangle \langle \psi|$ is the reduced density matrix of the subsystem A. The scatter plots, showing entanglement entropy of all eigenstates $|E_n\rangle$ as a function of their energy $E_n$, are displayed in Figs.~\ref{fig:spacing}(d), (e) and (f).  Here we take into account the translation symmetry of the system and work in the momentum sector $k=0$ for $H_1$ and $H_2$, and consider only the largest connected component and the inversion sector $I=0$ for $H_3$.
The results for other sectors are qualitatively similar.

Entanglement entropy distribution in Figs.~\ref{fig:spacing}(d) and (e) reveals a striking difference between the  Hamiltonians $H_1$ and $H_2$, even though they only differ by a free-boson hopping term, Eq.~(\ref{eq:h12}).   The model $H_1$ is constrained, which leads to a large spread of the entropy distribution and many low-entropy eigenstates including in the bulk of the spectrum. 
 From this perspective, $H_1$ is reminiscent of PXP model~\cite{ TurnerPRB,Khemani2018}. By contrast, $H_2$ has no such constraints and in this case the entanglement entropy is approximately a smooth function of the eigenstate energy. The Hamiltonian $H_3$ is doubly constrained, and this is reflected in its entanglement distribution, which also shows a large spread and several disconnected bands, reminiscent of an integrable system like the XY model~\cite{Alba2009}.

\subsection{Global quenches} 
 
The constraints in the models in Eqs.~\eqref{eq:h1}, \eqref{eq:h2} and \eqref{eq:h3} have significant effects on the dynamics governed by these Hamiltonians. We probe the dynamics by performing a global quench on the system. We assume the system is isolated and prepared in one of the Fock states, $|\psi_0\rangle$, at time $t=0$. We restrict to $|\psi_0\rangle$ being product states which are not necessarily translation-invariant, as such states are easier to prepare in experiment. However, our results remain qualitatively the same if we consider translation-invariant $|\psi_0\rangle$. After preparing the system in the state $|\psi_0\rangle$, which is not an eigenstate of the Hamiltonian, the system is let to evolve under unitary dynamics,
\begin{eqnarray}\label{eq:dyn}
|\psi(t)\rangle = \exp\left(-\frac{i}{\hbar} H t\right)|\psi_0\rangle.
\end{eqnarray}
where $H$ is one of the Hamiltonians of interest. From the time-evolved state, we evaluate the quantum fidelity,
\begin{eqnarray}\label{eq:fid}
F(t) = |\langle \psi_0 | \psi(t)\rangle |^2,
\end{eqnarray}
i.e., the probability for the wave function to return to the initial state. 
In a general many-body system, fidelity is expected to decay as $F(t)\sim \exp(-L(Jt)^2)$. It thus becomes exponentially suppressed in the system size 
for any fixed time $t^*$,
i.e., $F(t^*)\sim \exp(-cL)$, where $c$ is a constant. 
In scarred models, such as the Rydberg atom chain, 
fidelity at the first revival peak 
occurring at a time $T$
still decays exponentially, but exponentially slower,
i.e., $F(T)\sim \exp(-c'L)$, with $c' \ll c$. 
In Ref.~\onlinecite{TurnerPRB}, for a finite system with $L\lesssim 32$ atoms, the fidelity at the first revival can be as high as $\sim 70\%$, and several additional peaks at times $nT$ are also clearly visible. 

We first consider the Hamiltonian $H_1$. Several configurations exhibit periodic revivals of the fidelity $F(t)$, which can in some cases be higher than $90\%$. Most of these configurations involve a very dense cluster of bosons such as $\lvert...0N10...\rangle$. 
In contrast, a completely uniform configuration $\lvert...111...\rangle$ thermalizes very quickly.
Here we focus on periodically-reviving configurations with density being as uniform as possible. One family of such reviving configurations involves $n$ unit cells made of 3 lattice sites:
\begin{eqnarray}\label{eq:config}
|210210\ldots 210\rangle \equiv |(210)^n\rangle.
\end{eqnarray}
Time evolution of the fidelity for the initial state $|(210)^n\rangle$ for different system sizes $L=3n$ is shown in Fig.~\ref{fig:dyn021}(a). The initial state is assumed to be the product state, e.g., $\lvert\psi_0\rangle=|210\rangle$ for $L=3$. The frequency of the revivals in Fig.~\ref{fig:dyn021} is approximately the same for all system sizes. 
We emphasize that  similar results are obtained for a translation-symmetric initial state, e.g., $\lvert\psi_0\rangle=\frac{1}{\sqrt{3}}\left(\lvert210\rangle+\lvert021\rangle+\lvert102\rangle\right)$. Both cases converge in the large system limit, and the differences are only significant for $L=3$ when the revival frequency of the initial state with transition symmetry differs from the frequencies of other system sizes.
  
In Fig.~\ref{fig:dyn021}(b) we compare the fidelity for the initial state in Eq.~(\ref{eq:config}) when it is evolved by all three Hamiltonians in Eqs.~(\ref{eq:h1})-(\ref{eq:h3}). The initial state is fixed to be  $\lvert(210)^5\rangle$. We observe that the dynamics with $H_3$  has very prominent revivals; in fact as we will later show, these revivals are perfect and their  period  is approximately twice the revival period for $H_1$. In contrast, for $H_2$ the fidelity quickly drops to zero without any subsequent revivals. 

Finally, in Fig.~\ref{fig:dyn021}(c) we plot the time evolution of entanglement entropy. As expected from the fast decay of the fidelity, the entropy for $H_2$ rapidly saturates to its maximal value. Moreover, as expected from the perfect revivals in $H_3$, the entropy in that case oscillates around a constant value close to zero. 
For $H_1$, we observe a relatively slow growth of entropy, with oscillations superposed on top of that growth, again similar to PXP model~\cite{Turner2017}. For the initial state that is not translation-invariant, it is important how we cut the system, e.g., $|...210|210...\rangle$ versus $|...2102|10...\rangle$. In the first case, the entanglement entropy remains zero for $H_3$ because no particle can hop from one subsystem to the other, while in the second case the entropy oscillates around a constant value, which is the case in Fig.~\ref{fig:dyn021}(c). 

 \begin{figure}[htb]
 \includegraphics[width=0.485\textwidth]{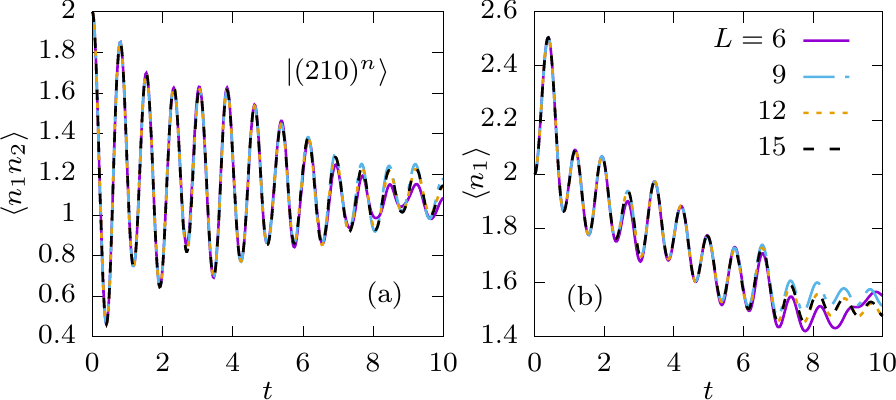}
\caption{Evolution of local observables for the Hamiltonian $H_1$. (a) Correlations between adjacent sites $\langle n_1n_2(t)\rangle$ for different system sizes and the initial state $|(210)^n\rangle$. (b) Density on one site $\langle n_1(t)\rangle$.
}\label{fig:correlations}
\end{figure}
In Fig.~\ref{fig:correlations} we show the $H_1$ evolution of two local observables, density correlations between two adjacent sites $\langle n_1n_2(t)\rangle$ and density on the first site $\langle n_1(t)\rangle$, starting from the initial state $|(210)^n\rangle$. Unlike fidelity and entanglement entropy, these observables can be easily measured in experiment. Both observables robustly oscillate with approximately the same frequency as the fidelity. The heights of the first few revival peaks are approximately converged for the system sizes ranging from $L=6$ to $L=15$, which suggests that revivals in such local observables can be observed in the thermodynamic limit. In the following Section, we will show that the oscillations observed in the dynamics from $|(210)^n\rangle$ state in Eq.~(\ref{eq:config}) and their frequency can be explained using a tractable model that involves only a small subset of all configurations in the Hilbert space, thus providing a realization of quantum scars in a correlated bosonic system.
Our starting point will be the model $H_3$, whose graph explicitly separates into disconnected subsets which makes the toy model exact, hence we can analytically calculate the revival frequency. Based on these results, we then introduce an approximation scheme that describes the dynamics from the same initial state under the $H_1$ Hamiltonian.

\subsection{Quantum scars in $H_1$ and $H_3$ models}\label{sec:h1h3}

The quench dynamics of fidelity and entanglement entropy in Fig.~\ref{fig:dyn021} suggest that $H_1$ and $H_3$ models are candidate hosts for many-body scarred eigenstates that can be probed by initializing the system in product states $|(210)^n\rangle$. We now analyze the structure of these states using our approach called ``cluster approximation" that is introduced in detail in Methods. 

The dynamics of $H_3$ within the sector containing the state $\lvert(210)^n\rangle$ 
can be solved exactly, as shown in Methods. 
The connected component of the state $\lvert(210)^n\rangle$ consists of all possible combinations of patterns $210$ and $120$. This means that triplets of sites evolve independently, and dynamically the system behaves as a collection of independent two level systems (spins-1/2). From this observation, it can be shown that revivals will be perfect with a period $T_3=\pi/2$. The same period is obtained for initial product state $\lvert(210)^n\rangle$ and its translation-invariant version; if the initial state is both translation-invariant and inversion-symmetric, the period is doubled. 
 
In contrast to the free dynamics in $H_3$, the $H_1$ model exhibits decaying revivals and does not admit an exact description. In order to approximate the quench dynamics and scarred eigenstates in $H_1$, we project the Hamiltonian to smaller subspaces of the full Hilbert space. These subspaces contain clusters of states which are poorly connected to the rest of the Hilbert space and thereby cause dynamical bottlenecks. As explained in Methods, the clusters can be progressively expanded to yield an increasingly accurate description of the dynamics from a given initial state.

For our initial state $|(210)^n\rangle$, the  minimal cluster is defined as one that contains all the states given by tensor products of  $210$, $120$ and $300$ patterns. 
Similar to the $H_3$ case, within this approximation, triplets of sites again evolve independently, and the dimension of the reduced Hilbert space is $ \mathrm{dim}\mathcal{H}^{\mathrm{c}} = 3^{L/3}$. 
The time-evolved state within the cluster is given by
\begin{eqnarray}\label{eq:e-h1}
 \lvert\psi^{\mathrm{c}}_n(t)\rangle=
\cos^n(4t)\lvert(210)^n\rangle+\ldots,
\end{eqnarray}
where the dots denote other configurations. The fidelity is
\begin{eqnarray}\label{eq:f-c1a}
 F^{\mathrm{c}}_{n}(t)=\lvert\langle\psi^{\mathrm{c}}_n(0)\lvert\psi^{\mathrm{c}}_n(t)\rangle\lvert^2=\lvert\cos(4t)\lvert^{2n}.
\end{eqnarray}
As in the case of $H_3$, this result is also valid for the translation-invariant  initial state. 
We see that the period of revivals is $T_1=\pi/4$, which is the same as for $H_3$ with a translation and inversion symmetric initial state. 

\begin{figure}[htb]
\includegraphics[width=0.4\textwidth]{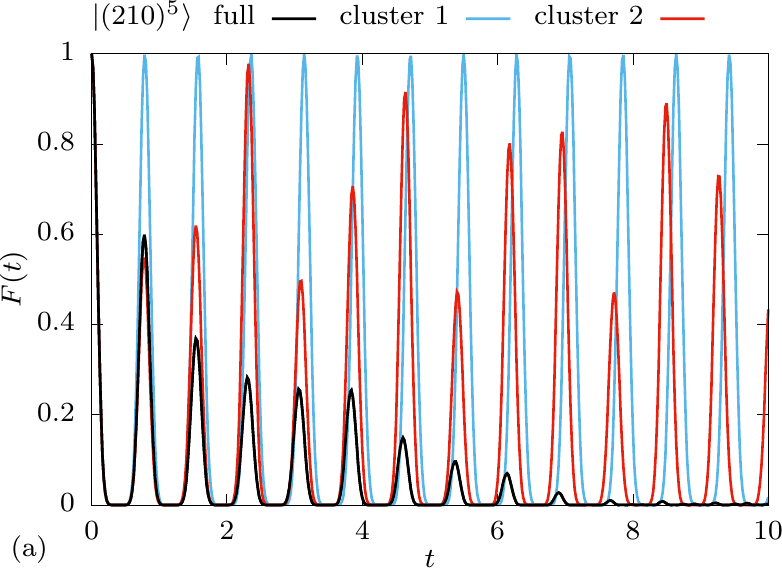}
\includegraphics[width=0.4\textwidth]{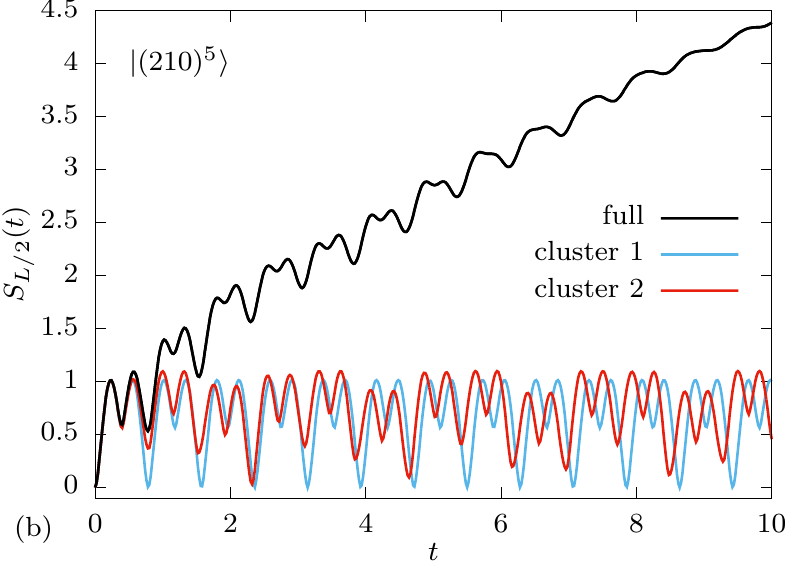}\\
\caption{Comparison of the full dynamics against the  minimal cluster ($1$) and extended cluster ($2$) approximation schemes. We consider the system size $L=15$ with the 
initial state $\lvert(210)^5\rangle$. 
(a) Time evolution of the fidelity. The frequency of revivals is approximately the same in both cases, but the results for the extended cluster show better agreement with the results for the full Hilbert space.
(b) Time evolution of the entanglement entropy. 
}\label{fig:cluster-dyn}
\end{figure}

The result of the cluster approximation is compared against the exact result for system size $L=15$ in Fig.~\ref{fig:cluster-dyn}. The frequency of the fidelity revival, shown by the blue line in Fig.~\ref{fig:cluster-dyn}(a), is accurately reproduced in this approximation, however the approximation does not capture the reduction in the magnitude of $F(t)$. Similarly, the  
dynamics of entanglement entropy, blue line in Fig.~\ref{fig:cluster-dyn}(b), is only captured at very short times. In particular, we observe that the maximum entanglement within the cluster remains bounded even at long times $t \sim 10$, while the exact entropy continues to increase and reaches values that are several times larger.

\begin{figure}[htb]
\includegraphics[width=0.45\textwidth]{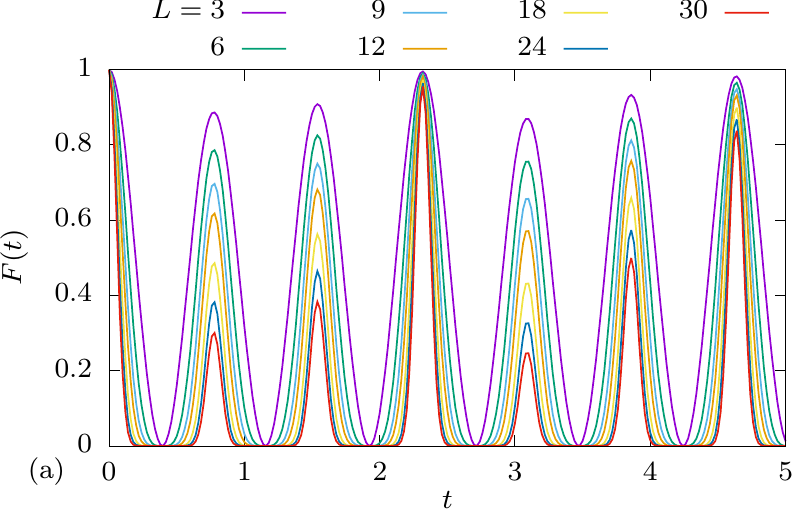}
\includegraphics[width=0.45\textwidth]{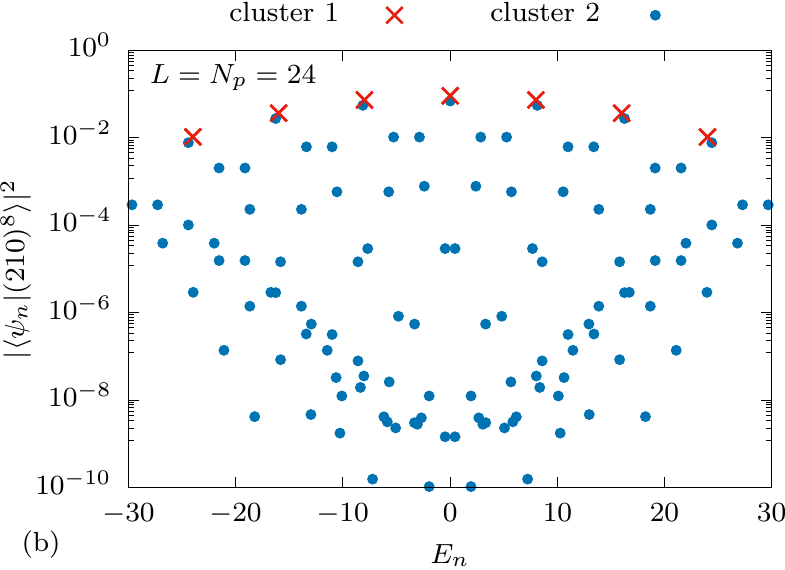}
\caption{ Cluster approximations. (a) Fidelity $F(t)$, for the Hamiltonian $H_1$ and initial states $\lvert(210)^n\rangle$, in the extended cluster approximation for various system sizes. 
(b) Eigenstate overlap with the initial state $\lvert(210)^8\rangle$ plotted on a log scale, for both cluster approximations. 
In the case of degenerate eigenstates the sum of their overlaps is shown.
}\label{fig:clusters}
\end{figure}

To obtain a more accurate approximation, we can expand the minimal cluster with several neighboring configurations on the graph. 
We define the extended cluster as a set of all states which can be obtained using tensor products of the configurations $210$, $120$, $300$ and $111$.
The enlarged cluster clearly contains the minimal cluster studied above, but it also includes additional configurations, resulting in a much better prediction for the first revival peak height, while still allowing for analytical treatment.
 The dimension of the extended cluster grows as
$ \mathrm{dim}\mathcal{H}^{\tilde{\mathrm{c}}} = 4^{L/3}$,
 and is thus exponentially larger than the minimal cluster approximation. Nevertheless, the extended cluster dimension is still exponentially smaller compared to the full Hilbert space, and within this approximation it is possible to numerically simulate the dynamics of larger systems, $L\lesssim 30$ -- see Fig.~\ref{fig:clusters}(a). The revivals are no longer perfect, while their frequency 
 is independent of the system size and 
 closer to the frequency of revivals for the full Hilbert space compared to to the minimal cluster approximation in Fig.~\ref{fig:cluster-dyn}.
 The overlap between the eigenstates of the Hamiltonian $H_1$ reduced to both the minimal and extended cluster and the state $\lvert(210)^8\rangle$ is given  in Fig.~\ref{fig:clusters}(b). 
The eigenstates that correspond to the minimal cluster approximately survive in the extended cluster, where they form a band with the highest overlap.

For the initial product state $(210)^n$, it is possible to analytically obtain the fidelity within the improved approximation for arbitrary system size. Similar to the previous methods, it can be shown (see Supplementary Note~2)
\begin{eqnarray}\label{eq:f-c2}
 F^{\tilde{\mathrm{c}}}_{L=3n}(t)&=&4^n\lvert b^2\cos(\alpha t)+d^2\cos(\beta t)\lvert^{2n},
 \end{eqnarray}
 where $\alpha=\sqrt{9+\sqrt{57}}\approx4.06815$, $\beta=\sqrt{9-\sqrt{57}}\approx1.20423$, $b\approx0.694113$ and $d\approx0.134933$.
Eq.~\eqref{eq:f-c2} is in excellent agreement with the numerical results in Fig.~\ref{fig:clusters}(a). 
It was also found to be a very good approximation for the translation-invariant initial state when $L\geq 9$ (data not shown).
 \begin{figure}[htb]
\includegraphics[width=0.45\textwidth]{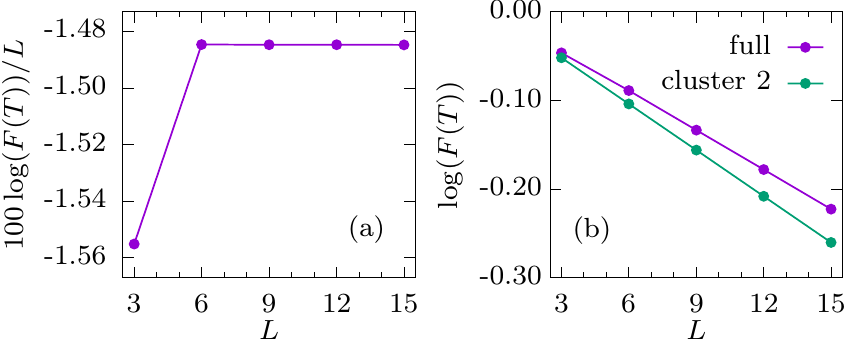}
\caption{ First peak height. (a) Logarithm (base 10) of the first revival peak divided by the system size, $\log(F(T))/L$, seems to saturate at a finite value in the thermodynamic limit. 
(b) Comparison of the logarithm of the first revival peak height for the full dynamics and the improved cluster approximation. The approximation serves as a lower bound.
}\label{fig:fph}
\end{figure}

Fig.~\ref{fig:fph}(a) shows that the logarithm of the fidelity per site, $\log(F(T))/L$, at the first peak, saturates at a finite value for large $L$.
 In the improved cluster approximation, the first peak height decays as 
 $\mathrm{e}^{-0.04L}$ (Supplementary Note~2). 
 For a completely random state, the fidelity would be $F\sim1/{\mathrm{dim}_\mathcal{H}}$. 
 In the case $\nu=1$ and large $L$, the Hilbert space dimension grows with the system size as
 \begin{equation}
 \mathrm{dim}_\mathcal{H}={{2L-1}\choose{L}}\approx{{2L}\choose{L}}\approx\frac{4^L}{\sqrt{\pi L}}.
 \end{equation}
 This back-of-the-envelope estimate suggests the fidelity of a random state is $F\sim\mathrm{e}^{-1.39L}$, which decays considerably faster than the first peak height in Fig.~\ref{fig:fph}. 
 The improved cluster approximation correctly reproduces the short-time dynamics, including the first revival peak, and sets a lower bound for the first peak height -- see Figs.~\ref{fig:cluster-dyn} and \ref{fig:fph}(b).

\begin{figure}[htb]
\includegraphics[width=0.4\textwidth]{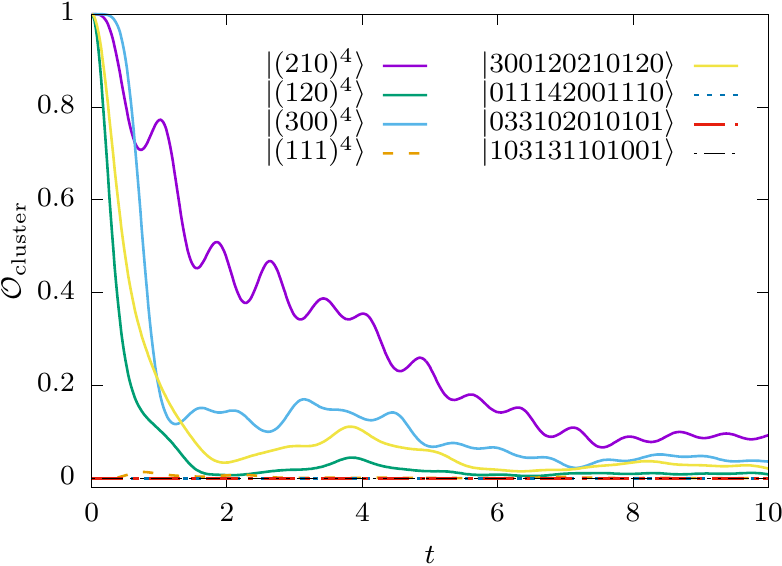}\\
\caption{Evolution of the probability to remain inside the minimal cluster. $\mathcal{O}_\mathrm{Cluster}$, as defined in Eq.~(\ref{eq:oc}). Initial configurations are indicated in the legend. Solid lines: configurations initially inside the cluster.
Dashed lines: configurations initially outside the cluster (all except $\lvert(111)^4\rangle$ are randomly chosen).
Similar results are obtained for the extended cluster (not shown). System size $L=12$.
}\label{fig:cluster-overlap}
\end{figure}
The evolution of the entanglement entropy for the extended cluster approximation is shown in Fig.~\ref{fig:cluster-dyn}(b). Inside the cluster,  entropy 
remains approximately constant with periodic oscillations that have the same frequency as the wave function revivals. Any further growth of the entanglement entropy can be attributed to the spreading of the wave-function outside the cluster.
To illustrate the ``leakage" of the wave function outside the cluster, in Fig.~\ref{fig:cluster-overlap} we compute the  time evolution of the overlap with a cluster, i.e., the probability to remain inside a cluster at time $t$, 
\begin{eqnarray}\label{eq:oc}
\mathcal{O}_\mathrm{Cluster} = \sum_{a \in \mathrm{Cluster}} |\langle a | \psi(t)\rangle|^2.
\end{eqnarray}
We consider several initial configurations that lie inside or outside the cluster. The configurations initially inside the cluster mostly stay there, and the configuration $\lvert(210)^4\rangle$ that has the highest revivals also has the highest overlap. Similarly, configurations initially outside the cluster continue to have negligible overlaps.  
The overlap starting from the configuration $\lvert(210)^4\rangle$ approximately predicts the revival peak heights for the full dynamics. 

 We now summarize the relation between $H_3$ and $H_1$ from the point of view of the cluster approximation. For the initial state $\lvert(210)^n\rangle$, the two models yield similar dynamics,  compare Eqs.~\eqref{eq:f-h3a}  and Eq.~\eqref{eq:f-c2}. The only difference is that the revival frequency is doubled in the latter case, which can be easily explained by the symmetry of the initial state and that of the Hamiltonian.  Hamiltonian $H_3$ is inversion-symmetric. If the initial state is also chosen to be inversion-symmetric, the frequency of the revivals doubles.
 The period is then $T_3^\mathrm{inv}=\pi/4$, which is  
 equal to the period of revivals $T_1$ of $H_1$ in the cluster approximation. This is also proven analytically in Methods, see Eq.~\eqref{eq:f-h3p}. For comparison, the revival period for the full Hilbert space is approximately $0.77$, which is slightly less than $\pi/4\approx0.79$. The Hamiltonian $H_1$ is not inversion-symmetric, so the frequency does not double for an inversion-symmetric initial state, but the revivals are lower in that case. 
 This shows that it is important for the symmetry of the initial state to match the symmetry of the Hamiltonian.

\begin{figure*}
\includegraphics[width=0.99\textwidth]{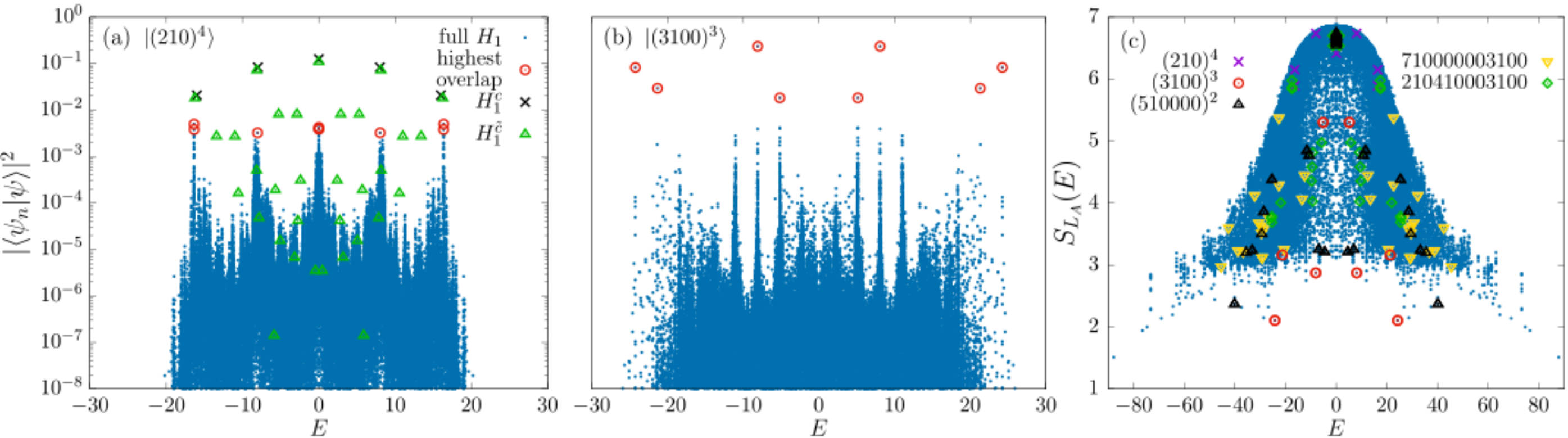}
\caption{Non-ergodic eigenstates.
(a) Overlap of the configuration $\lvert(210)^4\rangle$ with all the eigenstates of $H_1$, $H_1^{c}$ and $H_1^{\tilde{c}}$ versus the eigenstate energy for sector $k=0$ and system size $L=N_p=12$. 
(b) Same for $\lvert(3100)^3\rangle$.
(c) Entanglement entropy, eigenstates which have the highest overlap with some product states are marked in different colors.
}\label{fig:coefs}
\end{figure*}

Finally, the eigenstates of $H_1$, projected to the subspace of the minimal cluster approximation, form several degenerate bands whose eigenenergies are equally spaced in integer multiples of $4$.  Interestingly, some of these eigenstates approximately survive in the full $H_1$ model, and they are precisely the eigenstates that have the highest overlap with the configurations $\lvert(210)^n\rangle$, see Fig.~\ref{fig:coefs}(a).  In small system sizes, such as $L=6$, the surviving eigenstates are also the lowest entropy eigenstates in the middle of the spectrum, which is reminiscent of quantum scars in the PXP model~\cite{TurnerPRB}. In larger systems, e.g.,  $L=12$, the surviving eigenstates are slightly lower in entropy than their neighbors, but are far from being the lowest entropy eigenstates, as can be seen in Fig.~\ref{fig:coefs}. The lowest entropy eigenstates have high overlaps with  other configurations, such as $\lvert(3100)^3\rangle$, as shown in Figs.~\ref{fig:coefs}(b) and \ref{fig:coefs}(c). 
In the case of $\lvert(210)^n\rangle$, the eigenstates surviving in the full system belong to every other band of eigenstates in the cluster approximation and the number of the surviving eigenstates is $n+1$. For even system sizes this counting includes a zero-energy eigenstate.
 In Methods we discuss in more details the generalization of the cluster approximations to the states of the form $\lvert(N10...0)^n\rangle$, which were also found to have robust revivals and high overlaps with some low-entropy eigenstates.

\section{Discussion}\label{s:conclusions}

In this paper, we have introduced three models of bosons with ``soft'' kinetic constraints, i.e., density-dependent hopping.  We have demonstrated that some of these models exhibit similar phenomenology to other realizations of quantum many-body scars, for example the Rydberg atom system~\cite{Bernien2017}. We have studied quantum dynamics of these systems by performing global quenches from tensor-product initial states. We have shown that both the connectivity of the Hilbert space and the relative magnitude of the hopping coefficients have dramatic effects on the dynamics. For certain initial configurations, the constraints can lead to slow thermalization and revivals in the quantum fidelity. The revival frequency can be predicted by considering an exponentially reduced subset of the Hilbert space. 
For a family of initial configurations of the form $\lvert(210)^n\rangle$,
we have derived analytical expressions for the evolution of quantum fidelity within this approximation, which accurately capture the revival frequency obtained from exact numerical data. One notable difference between scarred dynamics in the present bosonic models and the PXP model is that the revivals exist in the absence of a hard kinetic constraint, i.e., in the fully connected Hilbert space. Our cluster approximation also explains the structure of some low-entropy eigenstates in the middle of the many-body spectrum. In addition, we have calculated the evolution of two local observables which are experimentally measurable, density correlations between two neighboring sites and density on a single site, and both of them show robust oscillations over a range of system sizes. We have also shown that the introduced models contain additional special properties, like the exponentially large zero-energy degeneracy which is related to the bipartite structure of the model.  

We now comment on the possible experimental realizations of the models we studied. The implementation of a correlated hopping term ($n_k b_i^{\dagger} b_j$) in optical lattices has attracted lot of 
attention due to a possible onset of quantum phases related to high-Tc  
superconductivity \cite{Eckholt2009}.
An early theoretical proposal exploits asymmetric interactions between 
the two atomic states in the presence of a state-dependent optical 
lattice \cite{Eckholt2009}. As a result, the obtained effective model 
corresponds to the inversion-symmetric form of $H_1$.
In addition, the same term has been found to feature as a higher-order 
correction of the standard Bose-Hubbard model~\cite{Mazzarella2006, 
Bissbort2012, Luhmann2012,  Dutta2015}. Although in this case the term 
typically represents a modification of the regular hopping term of the 
order of several percent, its contribution was directly measured~\cite{Jurgensen2014, Baier2016}. More  recently, the set of quantum models accessible in cold-atom 
experiments has been enriched through the technique of Floquet 
engineering~\cite{Eckardt2017}. As a notable example, a suitable driving 
scheme can renormalize or fully suppress the bare tunneling rate~\cite{Eckardt2009}. On top of that, by modulating local interactions an 
effective model with the density-dependent tunneling term has been 
engineered~\cite{Meinert2016}. For the models considered in this paper 
the most promising is a more recent driving scheme exploiting a double 
modulation of a local potential and on-site interactions~\cite{Zhao2019}. Related sophisticated driving schemes have already 
enabled a realization of dynamical gauge fields~\cite{Gorg2018, 
Barbiero2019, Schweizer2019} where both the amplitude and the phase of 
the effective tunneling are density-dependent.
Although these experimental proposals explain how to realize some of the correlated hopping terms present in our models using ultracold atoms in optical lattices, finding a scheme that exactly realizes our models requires further study. 
We emphasize that other models which would exhibit non-ergodic dynamics and scarred eigenstates as a result of the same mechanism that was explained in this work could be built, for example a linear combination of $H_1$ and $H_2$.

{\sl Note added:} During the completion of this work, we became  aware of Ref.~\onlinecite{Pancotti2019} which identified non-thermal eigenstates and slow dynamics in the quantum East model. 
Moreover, a recent study~\cite{Zhao2020} proposed a Floquet scheme for 
a bosonic model with density-assisted hopping,
finding signatures of quantum many-body scars.

\section{Methods}\label{sec:methods}

In order to more efficiently describe the dynamics of our models, we introduce a method -- ``cluster approximation", that is based on Hilbert space truncation inspired by the bipartite 
graph structure of $H_1$.  Before providing details about the cluster approximation for $H_1$ and its generalizations, we present an exact solution for the perfect revivals in $H_3$ model, which serves as a motivation for the more complicated case of $H_1$.

\subsection{Bipartite lattice and zero modes}\label{sec:bipartite}

 The graph of $H_1$ is bipartite, i.e. all the basis configurations can be divided into two disjoint sets, and the action of the Hamiltonian connects configurations in one set only to the configurations in the other and vice-versa (the Hamiltonian is off-diagonal).
  One way to sort configurations into these two sets is by parity of the quantity
 \begin{eqnarray}\label{eq:dist2}
 \Delta_a=\frac{\abs{n_\mathrm{even}-n_\mathrm{odd}+C}}{2},
 \end{eqnarray}
 where $C=0$ if $L$ is even and $C=1$ if $L$ is odd.
We define
$n_\mathrm{even}$ and $n_\mathrm{odd}$ as the total numbers of particles at even and odd sites, respectively,
 \begin{eqnarray}\label{eq:even_odd}
 n_\mathrm{even}=\sum_{l=1}^{L_1}n_{2l},\ n_\mathrm{odd}=\sum_{l=1}^{L_2}n_{2l-1},
 \end{eqnarray}
 where $L_1=L_2=L/2$ if $L$ is even, and $L_1=(L-1)/2$, $L_2=(L+1)/2$ if $L$ is odd.
 If only nearest neighbor hoppings are allowed and if no two odd sites are coupled (if the system has open boundary conditions for any $L$ or periodic boundary conditions for $L$-even), each hopping either increases $n_\mathrm{even}$ by one and decreases $n_\mathrm{odd}$ by one, or vice-versa. This means that each hopping can change $\Delta_a$ only by $\pm1$.
 
  In special cases, like $H_1$ at filling factor $\nu=1$, it is also possible to define quantities like $\Delta_a$ for odd system sizes and periodic boundary conditions.  This is a consequence of the constraints imposed by $H_1$, i.e., the fact that a particle cannot hop to an empty site to its left  (Supplementary Note~3).
  Note that $H_2$ in the same geometry is not bipartite.
 
 Another way to sort configurations into two sets is by parity of the distance from the configuration $\lvert111...111\rangle$, which we define as
 \begin{eqnarray}\label{eq:dist}
 d_a=\mathrm{min}_n \{ \langle 111...111\lvert H_1^n\lvert a\rangle \neq 0 \}. 
 \end{eqnarray}
 In this case, the two sets are the configurations with even and with odd distances $d_a$. 
 One hopping can change $d_a$ only by $\pm1$ or $0$.
 Changes by other values are not possible by definition if the Hamiltonian is Hermitian (all hoppings are reversible).
 Both $d_a$ and $\Delta_a$ have the same parity, thus $d_a$ must always change after one hopping in even system sizes or in systems with open boundary conditions.
 As a consequence, $d_a$ cannot change by $0$ if $\Delta_a$ can only change by $\pm1$.

 The graphs of bipartite systems do not contain any loops of odd dimension (triangles, pentagons, heptagons and so on). Moreover, the energy spectra of bipartite systems are symmetric around zero. Their Hamiltonians anticommute with the operator $(-1)^{\Delta_a}$. 
 The presence of such an operator in a bipartite lattice leads to exact zero energy states in the spectrum~\cite{Sutherland86,Inui}. 
 It can be shown that the exponentially growing number of zero modes of $H_1$ is related to the difference between the numbers of elements in the two sets of its bipartite graph (Supplementary Note~4). 
 Additionally, the algebraic structure of zero energy eigenstates can be explained by the structure of the graph -- such eigenstates can be constructed as superpositions of configurations from only one of the sets.
 Similar properties are found for $H_2$ for even $L$, as its graph is also bipartite in that case. The properties of the zero-energy manifold are discussed in more detail in Supplementary Note~4.

\subsection{Perfect revivals in the $H_3$ model}\label{sec:h3}

We start with a warmup calculation for $H_3$ acting on $L=3$ sites. The connected subspace of $210$ contains only two configurations, $120$ and $210$. The Hamiltonian reduced to this subspace is
\begin{eqnarray}
H^{'}_3=-
\begin{pmatrix}
0 & 2\\
2 & 0
\end{pmatrix},
\end{eqnarray}
where the basis vectors are
\begin{eqnarray}
\begin{pmatrix}
1 \\
0
\end{pmatrix}
= \lvert210\rangle,\ 
\begin{pmatrix}
0 \\
1
\end{pmatrix}
= \lvert120\rangle.
\end{eqnarray}
The eigenvalues of $H_3^{'}$ are $E_1=-2$ and $E_2=2$.
The initial state $\lvert\psi_1(t=0)\rangle=\lvert210\rangle$ evolves as
\begin{eqnarray}
 \lvert\psi_1(t)\rangle=
\cos(2t)\lvert210\rangle-i\sin(2t)\lvert120\rangle,\label{eq:dimH3L3}
\end{eqnarray}
and the state $\lvert\psi_2(t=0)\rangle=\lvert120\rangle$ evolves as
\begin{eqnarray}
 \lvert\psi_2(t)\rangle=
-i\sin(2t)\lvert210\rangle+\cos(2t)\lvert120\rangle.
\end{eqnarray}

Previous results can be straightforwardly generalized to larger systems. Let the length of the system be $L=3n$ for simplicity. The connected component of the state $\lvert(210)^n\rangle$ consists only of 
combinations of patterns $210$ and $120$, which means that triplets of sites evolve independently. 
From Eq.~\eqref{eq:dimH3L3}, the initial state $\lvert\psi_n(t=0)\rangle=\lvert(210)^n\rangle$ evolves as
\begin{eqnarray}
\nonumber \lvert\psi_{L=3n}(t)\rangle &=& \cos^n(2t)\lvert(210)^n\rangle \\
&+& (-i)^n \sin^n(2t)\lvert(120)^n\rangle + ... 
\end{eqnarray}
where  ``$...$'' denotes contributions of the basis configurations other than $\lvert(210)^n\rangle$ or $\lvert(120)^n\rangle$. The fidelity is
\begin{eqnarray}\label{eq:f-h3a}
 F_{L=3n}(t)=\lvert\langle\psi_n(0)\lvert\psi_n(t)\rangle\lvert^2=\lvert\cos2t\lvert^{2n}.
\end{eqnarray}
It follows that the revivals are perfect, with a period $T_3=\pi/2$.
This result is also valid for the translation-invariant initial state $\lvert(210)^n\rangle_\mathrm{T}$,
\begin{eqnarray}\label{eq:transl-inv}
\lvert(210)^n\rangle_\mathrm{T} \equiv \frac{1}{\sqrt{3}}\left(\lvert(210)^n\rangle+\lvert(021)^n\rangle+\lvert(102)^n\rangle\right),\quad\quad
\end{eqnarray}
as the connected subspaces of $210$, $021$ and $102$ do not overlap and therefore evolve independently. 

However, an initial state that is both translation symmetric and inversion symmetric has different dynamics.
The inverse of the configuration $\lvert(210)^n\rangle$ is the configuration $\lvert(012)^n\rangle$, which is a translation of the state $\lvert(120)^n\rangle$ that belongs to the connected subspace of $\lvert(210)^n\rangle$.
The initial state 
\begin{eqnarray}
\lvert\psi^\mathrm{inv}_n(t=0)\rangle=\frac{1}{\sqrt{2}}\lvert(210)^n\rangle_\mathrm{T}+\frac{1}{\sqrt{2}}\lvert(120)^n\rangle_\mathrm{T}
\end{eqnarray}
evolves as
\begin{eqnarray}
\lvert\psi^\mathrm{inv}_n(t)\rangle = 
\left(\cos^n2t+(-i)^n\sin^n2t\right)\lvert\psi^\mathrm{inv}_n(t=0)\rangle
+...\quad\quad
\end{eqnarray}
and the fidelity is
\begin{eqnarray}\label{eq:f-h3p}
 F^\mathrm{inv}_{n}(t) &=&\lvert\langle\psi^\mathrm{inv}_n(0)\lvert\psi^\mathrm{inv}_n(t)\rangle\lvert^2\nonumber\\
&=&\lvert\cos^n2t+(-i)^n\sin^n2t\lvert^{2}.
\end{eqnarray}
The frequency of the revivals is now doubled, so the period is $T^\mathrm{inv}_3=\pi/4$.

\subsection{Cluster approximations for the $H_1$ model}\label{sec:h1_clusters}

\begin{figure}[htb]
\includegraphics[width=0.45\textwidth]{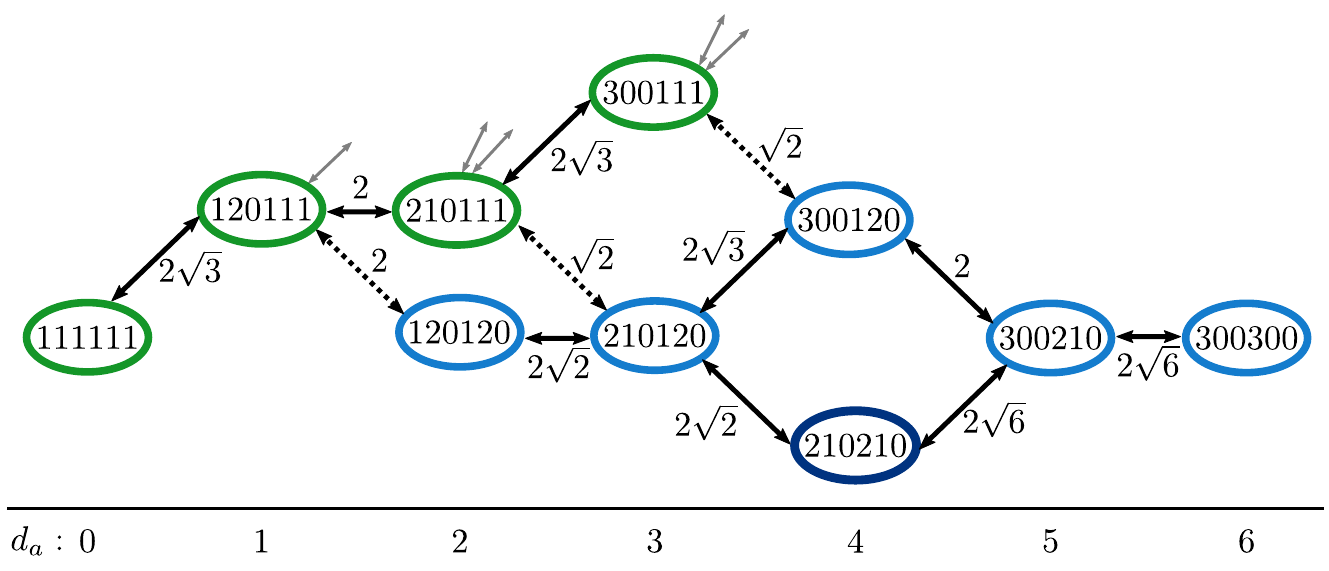}
\caption{Minimal and extended clusters.
Hamiltonian $H_1$ and system size $L=N_\mathrm{p}=6$. Configurations labelled inside the ellipses are representatives of an orbit of translation symmetry.
The minimal cluster is defined by the blue configurations, while green configurations represent the additional components of the extended cluster.
Grey arrows connect to configurations outside the extended cluster.
The numbers bellow the graph show the distance $d_a$ from the configuration $111111$ evaluated using Eq.~(\ref{eq:dist}).
}\label{fig:021021}
\end{figure}

 Here we introduce a scheme for approximating the dynamics from initial states $(210)^n$ in the $H_1$ model.
 As can be observed in Fig.~\ref{fig:dyn021}, the revival periods are approximately the same for different system sizes. We first focus on the non-trivial case $L=6$. Fig.~\ref{fig:021021} shows part of the graph that contains the initial state, $\lvert210210\rangle$. Configurations labelled inside the ellipses denote representatives of an orbit of translation symmetry, i.e., the configurations are translation-invariant such as the one in Eq.~\eqref{eq:transl-inv}.  

The minimal subcluster of the graph is highlighted in blue color in Fig.~\ref{fig:021021}. This cluster  is indeed weakly connected to the rest of the configuration space, as it has only $3$ connections that lead outside this cluster (dashed lines) and their hopping coefficients are slightly lower in magnitude than those inside the cluster, meaning that the probability is higher to stay inside the cluster than to leave.  
 The hopping coefficients leading outside are not significantly smaller than the coefficients staying inside, but in combination with the relatively small number of connections this has significant effects on the dynamics.
 This effect is even more pronounced when the difference in magnitudes is further increased by squaring the particle number operators (see Supplementary Note~1). 
  
 The minimal cluster from Fig.~\ref{fig:021021} contains all the states given by tensor products of  $210$, $120$ and $300$ configurations.
 The set of configurations belonging to this cluster could have been chosen differently, but this particular choice has at least two advantages.
 Firstly,
 inside this cluster, the evolution of the configuration $\lvert210210\rangle$ can be thought of as two subsystems $210$ evolving separately. The evolution of all such configurations at different system sizes can be reduced to the evolution of $L=3$ subsystems $210$, similar to the case of $H_3$ in the connected subspace of $(210)^n$. 
 Secondly, this definition allows easy generalization to different system sizes $L=3n$ with initial states $(210)^n$.
 We would like to emphasize that the cluster was not chosen arbitrarily.
 The calculations of the probability density distribution starting from the initial configuration $\lvert210210\rangle$ and evolving with $H_1$ have shown that the probability density stays high in this region of the Hilbert space as long as the revivals in fidelity are visible.
 The configurations important for the dynamics were then identified by analyzing the structure of the graph around the initial configuration.

As an example, consider system size $L=3$.  The reduced Hilbert space of the cluster $\mathcal{H}^{\mathrm{c}}$ is spanned by the (non-translation-invariant) configurations
\begin{eqnarray}
\begin{pmatrix}
1 \\
0 \\
0
\end{pmatrix}
= \lvert300\rangle,\
\begin{pmatrix}
0 \\
1 \\
0
\end{pmatrix}
= \lvert210\rangle,\ 
\begin{pmatrix}
0 \\
0 \\
1
\end{pmatrix}
= \lvert120\rangle.
\end{eqnarray}
The Hamiltonian reduced to this subspace is
\begin{eqnarray}
H_{1}^{\mathrm{c}}=-
\begin{pmatrix}
0 & 2\sqrt{3} & 0\\
2\sqrt{3} & 0 & 2\\
0 & 2 & 0
\end{pmatrix},
\end{eqnarray}
and its eigenvalues are $E_1=-4$, $E_2=4$, $E_3=0$.
The initial configuration $\lvert210\rangle$ evolves according to 
\begin{eqnarray}
 \lvert\psi^{\mathrm{c}}_1(t)\rangle&=&
-\frac{i}{2}\sin(4t)\left(\sqrt{3}\lvert300\rangle+\lvert120\rangle\right)\nonumber\\&&+\cos(4t)\lvert210\rangle.
\end{eqnarray}
By generalizing this result to larger systems, it is easy to prove Eqs.~\eqref{eq:e-h1} and \eqref{eq:f-c1a}.

The minimal clusters can be expanded by adding several neighboring configurations.
For similar reasons as in the case of minimal clusters, the extended clusters are defined as sets of all states which can be obtained using tensor products of the configurations $210$, $120$, $300$ and $111$.
In the case of $L=6$, the enlarged cluster can be observed in Fig.~\ref{fig:021021}.
It contains the minimal cluster studied previously, but it also includes additional configurations shown in green ellipses.
Again, the approximation could be improved by including more configurations, but this particular choice is well suited for analytical treatment (Supplementary Note~2) and, as shown above,  it gives a good prediction for the first revival peak height.

\subsection{Generalization to other clusters}\label{sec:generalization}

Building on the previous observation that some of the low-entropy eigenstates have large weight on $\lvert(3100)^3\rangle$ product state, we have investigated periodic revivals from such a larger class of initial states. We find that robust revivals are associated with initial product states of the form
\begin{eqnarray}\label{eq:conf}
\lvert((N-1)1\underbrace{0...0}_{N-2})^n\rangle,
\end{eqnarray}
where $N$ is the length of the unit cell ($L=N n$). For example, some of these configurations are $\lvert(3100)^n\rangle$, $\lvert(41000)^n\rangle$ and $\lvert(510000)^n\rangle$. Combinations of those patterns such as $\lvert310041000\rangle$ also exhibit similar properties, but we will restrict ourselves to the simpler former cases.

We can construct a generalization of the cluster approximation for configurations of the form in Eq.~\eqref{eq:conf}. As in the case of $\lvert(210)^n\rangle$, the dynamics inside one unit cell explains the dynamics of the full system. 
The generalized clusters can be chosen in such a way that their Hilbert spaces are spanned by $N$ configurations
\begin{eqnarray}\label{eq:cluster_basis}
\lvert i\rangle=\lvert((N+1-i)(i-1)\underbrace{0...0}_{N-2})^n\rangle,
\end{eqnarray}
where $i$ takes values $1,2,\ldots N$. If we consider only one unit cell ($n=1$), the graph that connects these configurations has a linear structure without any loops, i.e., each configuration $\lvert i \rangle$ is solely connected to the configurations $\lvert i \pm 1\rangle$, except the two configurations at the edges, $\lvert 1 \rangle$ and $\lvert N \rangle$, which are only connected to $\lvert 2 \rangle$ and $\lvert N-1 \rangle$, respectively.

The projection of the Hamiltonian $H_1$ to this cluster, which we denote by $H_1^\mathrm{c}$, has a very simple structure: it has the form of a tight-binding chain with the only nonzero matrix elements on the upper and lower diagonals:
\begin{eqnarray}\label{eq:matrix_elements}
H_{1;i,i+1}^\mathrm{c}=H_{1;i+1,i}^\mathrm{c}=(N-i)\sqrt{i(N+1-i)}.
\end{eqnarray}
The dynamics within a single unit cell under $H_1^\mathrm{c}$ corresponds to density fluctuations between the first and the second site. 
Following the same procedure as previously, we can now diagonalize $H_1^\mathrm{c}$ and compute the fidelity time series for the initial configuration $\lvert(N-1)10...0\rangle$. 
This result can be directly generalized to configurations of the form $\lvert((N-1)10...0)^n\rangle$. 
The derivation is valid for both translation-invariant and non-translation-invariant initial configurations, as the cluster in Eq.~\eqref{eq:cluster_basis} is disconnected from its translated copies. We stress that this disconnection, namely the absence of a hopping term between $\lvert 1(N-1)0...0\rangle$ and $\lvert 0N0...0\rangle$, is a consequence of the constraints imposed by $H_1$ and it would not hold for $H_2$. In this way, we have calculated the time evolution of the fidelity starting from the configurations $\lvert(3100)^n\rangle$ (for $n=1,2,3,4$), $\lvert(41000)^n\rangle$ ($n=1,2,3$) and $\lvert(510000)^n\rangle$ ($n=1,2$), and compared it with the exact numerical results for the full $H_1$. The cluster approximation captures both the revival frequency and the height of the first peak. Similar to the $\lvert(210)^n\rangle$ case, the approximation can be improved by adding further configurations to the clusters. Moreover, if we want to consider translation-invariant initial  states, we can follow the same recipe for $\lvert(210)^n\rangle$ by summing translated patterns with the required phase factors given in terms of momenta in multiples of $2\pi/N$. We have checked that revivals appear in these momentum sectors, with roughly the same frequency.

We note that the configurations with larger units cells thermalize more quickly on shorter timescales, but slower at long times. 
Initially, the states starting from configurations with smaller $N$ have lower entanglement entropies than those with larger $N$. 
The Hilbert spaces of large $N$ unit cells are larger, so the entanglement entropy starting from these configurations rapidly grows to the maximal value for that unit cell. 
However, the only way for the wave-function to spread through the entire Hilbert space is that a unit cell reaches a state close to $111...111$, so that particles can hop to the other unit cells. This is less likely for large $N$, and therefore such configurations need long times to fully thermalize.
As a result, smaller $N$ entanglement entropies grow faster and after long enough time they overtake those for larger $N$. 
For example, in the case of $L=12$ and translation-invariant initial states, $(210)^4$ overtakes $(3100)^3$ and $(510000)^2$ around $t\sim2$, and $(3100)^3$ overtakes $(510000)^2$ around $t\sim80$  (Supplementary Fig.~3). 

Finally, we note that non-thermal behavior reminiscent of  the one studied here was previously observed in an unconstrained Bose-Hubbard model, for example in the context of ``arrested expansion"~\cite{Arrested1,Arrested2} and quenches from superfluid to Mott insulator phase~\cite{Greiner02,Kollath2007}. In these cases, the main ingredient is the strong on-site interaction, which causes the energy spectrum to split into several bands. 
Due to the large energy differences between bands, the dynamics of an initial state from a particular band is at first limited only to the eigenstates that belong to the same band. Additionally, these energy bands are approximately equally spaced, which can lead to revivals in fidelity if several bands are populated. 
In contrast, our models do not feature on-site interaction, and  the mechanism which slows down the spread of the wave function is correlated hopping, which suppresses connections between certain configurations and modifies the hopping amplitudes between others, thus creating bottlenecks that separate different clusters of states.

\section{Data Availability}
The data that support the plots within this paper and other findings of this study are available at
\href{https://doi.org/10.5518/810}{https://doi.org/10.5518/810}~\cite{Data}.

\section{Code Availability}
Code is available upon reasonable request.

\section{Acknowledgments}

The authors thank Thomas Iadecola for fruitful discussions.
A.H. and I.V. acknowledge funding provided by the Institute of Physics Belgrade, through the grant by the Ministry of Education, Science, and Technological Development of the Republic of Serbia.
N.R. was supported by the Department of Energy Grant No. DE-SC0016239, the National Science Foundation EAGER Grant No. DMR 1643312, Simons Investigator Grant No. 404513, ONR Grant No. N00014-14-1-0330, the Packard Foundation, the Schmidt Fund for Innovative Research, and a Guggenheim Fellowship from the John Simon Guggenheim Memorial Foundation.
Z.P. acknowledges support by EPSRC grant EP/R020612/1 and the National Science Foundation under Grant No. NSF PHY-1748958.
Part of the numerical simulations were performed on the PARADOX-IV supercomputing facility at the Scientific Computing Laboratory, National Center of Excellence for the Study of Complex Systems, Institute of Physics Belgrade. 
The authors would also like to acknowledge the contribution of the COST Action CA16221.‌

\section{Author contributions}

A.H., I.V., N.R. and Z.P. contributed to developing the ideas,
analyzing the results and writing the manuscript. A.H. performed the calculations and designed the figures. 

\section{Competing interests}
The authors declare no competing interests.

%

\end{document}


\title{Supplementary Information}
 \author{Ana Hudomal}
\author{Ivana Vasi\'c}
\author{Nicolas Regnault}
\author{Zlatko Papi\'c}


\thispagestyle{empty}
\begin{center}
 \Large {\bf Supplementary Information}
\end{center}

\section*{Supplementary Figures}

\begin{figure*}[htb!]
\includegraphics[width=0.4\textwidth]{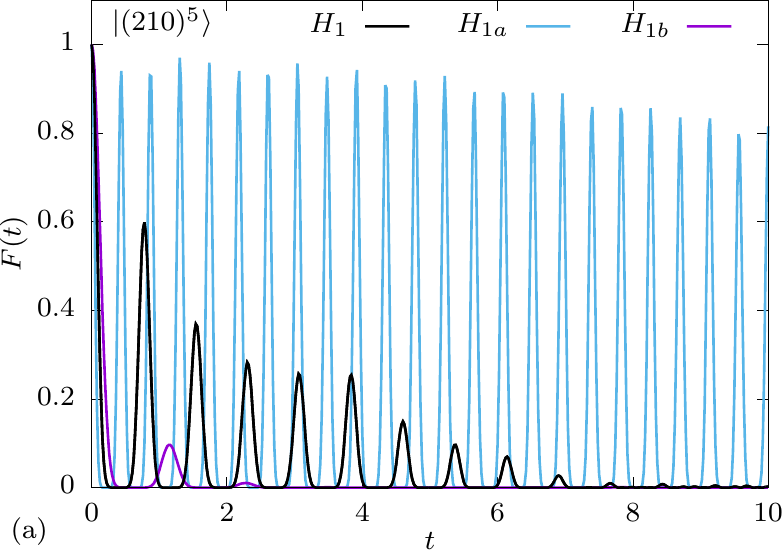}
\includegraphics[width=0.4\textwidth]{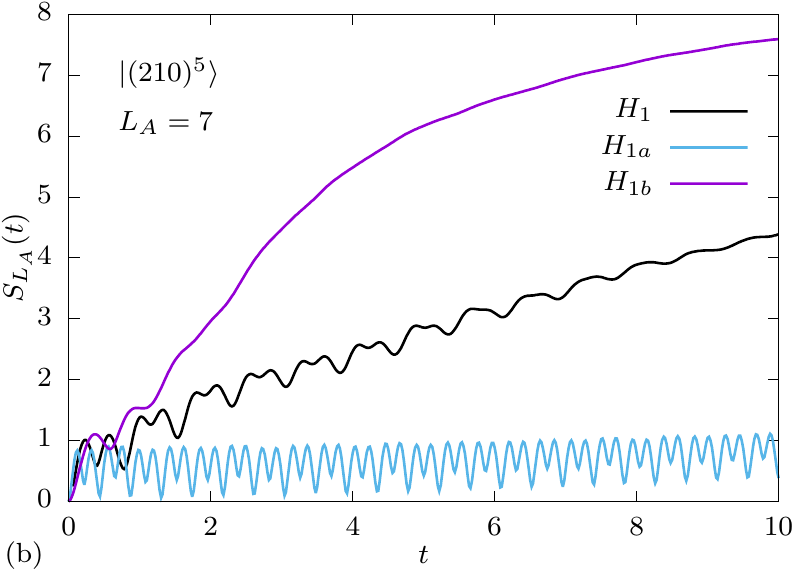}
\caption{Time evolution of (a) fidelity $\lvert \langle \psi(t)\lvert\psi(0)\rangle\lvert^2$ and (b) entanglement entropy $S_{L_\mathrm{A}=7}(t)$ governed by three different Hamiltonians, $H_1$, $H_{1\mathrm{a}}$ and $H_{1\mathrm{b}}$. System size $L=15$ and  initial state $\lvert(210)^5\rangle$.
}\label{fig:dynH1ab}
\end{figure*}

\begin{figure}[htb]
\includegraphics[width=0.6\textwidth]{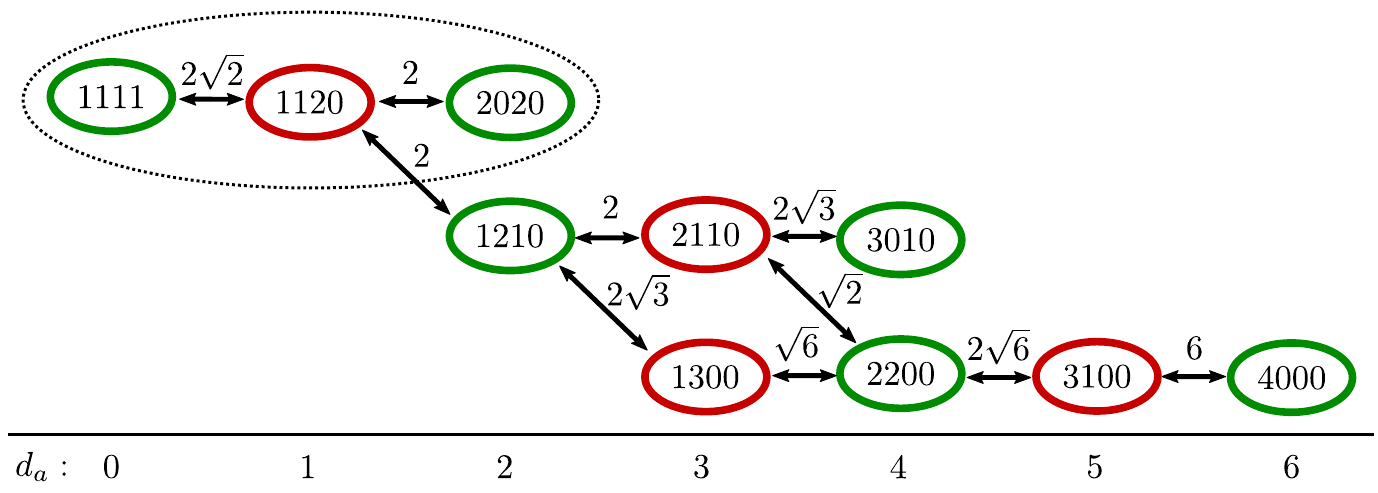}
\caption{Bipartite graph for $H_1$ and $L=N_\mathrm{p}=4$. The two classes of configurations are shown in green (even) and red (odd) ellipses. The configurations are written in the translation-invariant basis. The arrows represent the action of the Hamiltonian $H_1$ and the numbers above the arrows are the magnitudes of the corresponding hopping coefficients. 
The numbers bellow the graph show the distance $d_a$ from the configuration $1111$.
}\label{fig:L4}
\end{figure}

\begin{figure*}[thb!]
\includegraphics[width=0.4\textwidth]{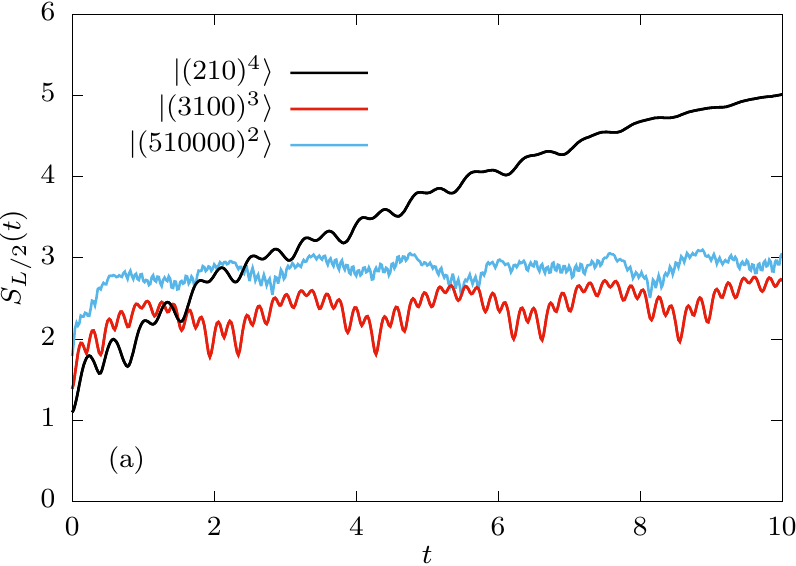}
\includegraphics[width=0.4\textwidth]{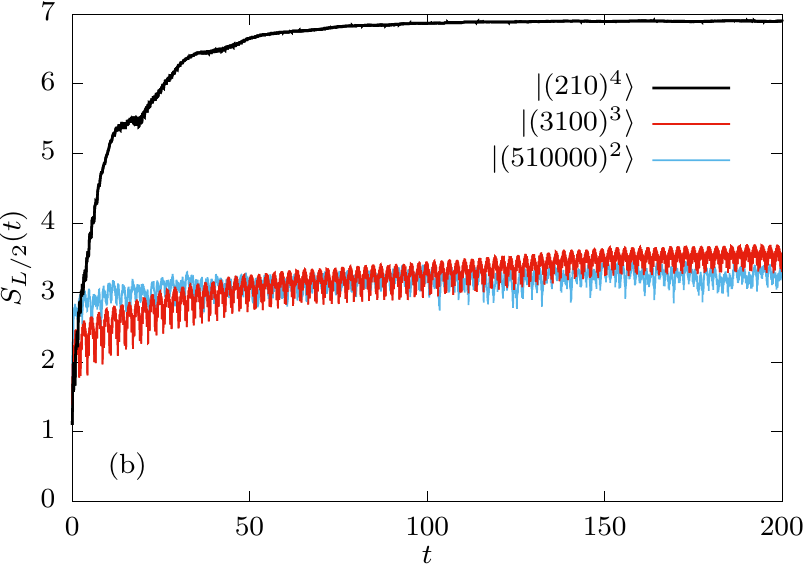}
\caption{Time evolution of entanglement entropy for three different translation-invariant initial states which exhibit slow thermalization.
(a) Short timescale. (b) Long timescale. 
System size $L=12$.
Configurations with larger unit cells (such as $\lvert(510000)^2\rangle$), thermalize more quickly than those with smaller unit cells (such as $\lvert(210)^4\rangle$) on shorter timescales, but slower at longer times.
}\label{fig:cluster_size}
\end{figure*}

\FloatBarrier

\section*{Supplementary Tables}

 
\begin{table}[htb]
\begin{tabular}{| c || c | c | c | c | c | c | c | c | c | c | c || c |}
\hline
 & \multicolumn{11}{c||}{$\frac{kL}{2\pi}$} &\\
 \hline 
 $L=N_\mathrm{p}$ & $0$ & $1$ & $2$ & $3$ & $4$ & $5$ & $6$ & $7$ & $8$ & $9$ & $10$ & total \\
 \hline 
 \hline 
 $2$ & $0$ & $1$ & & & & & & & & & & $1$\\
 \hline 
 $3$ & $0$ & $1$ & $1$ & & & & & & & & & $2$\\
 \hline 
 $4$ & $2$ & $0$ & $1$ & $0$ & & & & & & & & $3$\\
\hline 
 $5$ & $2$ & $1$ & $1$ & $1$ & $1$ & & & & & & & $6$\\
\hline
 $6$ & $2$ & $3$ & $0$ & $4$ & $0$ & $3$ & & & & & & $12$\\ 
\hline
 $7$ & $2$ & $3$ & $3$ & $3$ & $3$ & $3$ & $3$ & & & & & $20$\\
 \hline
 $8$ & $10$ & $0$ & $8$ & $0$ & $9$ & $0$ & $8$ & $0$ & & & & $35$\\
 \hline
 $9$ & $8$ & $8$ & $8$ & $7$ & $8$ & $8$ & $7$ & $8$ & $8$ & & & $70$\\
 \hline
 $10$ & $4$ & $25$ & $2$ & $25$ & $2$ & $26$ & $2$ & $25$ & $2$ & $25$ & & $138$\\
 \hline
 $11$ & $22$ & $23$ & $23$ & $23$ & $23$ & $23$ & $23$ & $23$ & $23$ & $23$ & $23$ & $252$\\
\hline
\end{tabular}
\caption{The number of zero-energy states for the Hamiltonian $H_1$ and different system sizes. The number of states is resolved per momentum sectors, denoted by their momentum indices $i$ that parametrize the  momenta $k_i=\frac{2\pi}{L}i$.
\label{tab:zm}}
\end{table}

\begin{table}[htb]
\begin{tabular}{| c || c | c | c | c | c | c | c | c | c | c | c || c |}
\hline
 & \multicolumn{11}{c||}{$\frac{kL}{2\pi}$} &\\
 \hline 
 $L=N_\mathrm{p}$ & $0$ & $1$ & $2$ & $3$ & $4$ & $5$ & $6$ & $7$ & $8$ & $9$ & $10$ & total \\
 \hline 
 \hline 
 $2$ & $0$ & $1$ & & & & & & & & & & $1$\\
 \hline 
 $3$ & $0$ & $0$ & $0$ & & & & & & & & & $0$\\
 \hline 
 $4$ & $2$ & $0$ & $1$ & $0$ & & & & & & & & $3$\\
 \hline 
 $5$ & $0$ & $0$ & $0$ & $0$ & $0$ & & & & & & & $0$\\
\hline
 $6$ & $0$ & $3$ & $0$ & $4$ & $0$ & $3$ & & & & & & $10$\\ 
\hline
 $7$ & $0$ & $0$ & $0$ & $0$ & $0$ & $0$ & $0$ & & & & & $0$\\
 \hline
 $8$ & $10$ & $0$ & $8$ & $0$ & $9$ & $0$ & $8$ & $0$ & & & & $35$\\
 \hline
 $9$ & $0$ & $0$ & $0$ & $0$ & $0$ & $0$ & $0$ & $0$ & $0$ & & & $0$\\
 \hline
 $10$ & $0$ & $25$ & $0$ & $26$ & $0$ & $26$ & $0$ & $26$ & $0$ & $25$ & & $128$\\
 \hline
 $11$ & $0$ & $0$ & $0$ & $0$ & $0$ & $0$ & $0$ & $0$ & $0$ & $0$ & $0$ & $0$\\
\hline
\end{tabular}
\caption{The number of zero-energy states for the Hamiltonian $H_2$ and different system sizes. The columns are the same as in Supplementary Table~\ref{tab:zm}. 
\label{tab:zm2}}
\end{table}


\begin{table}[htb]
\begin{tabular}{| c | c | c | c | c |}
\hline 
 {} & \multicolumn{2}{c|}{all sectors} & \multicolumn{2}{c|}{$k=0$}\\
\hline
 $L=N_\mathrm{p}$ & $g-r$ & $N_0$ & $g-r$ & $N_0$\\
 \hline 
 $2$ & $-1$ & $1$ & $0$ & $0$\\
\hline 
 $3$ & $-2$ & $2$ & $0$ & $0$\\
 \hline 
 $4$ & $3$ & $3$ & $2$ & $2$\\
 \hline 
 $5$ & $6$ & $6$ & $2$ & $2$\\
\hline
 $6$ & $-10$ & $12$ & $0$ & $2$\\ 
\hline
 $7$ & $-20$ & $20$ & $-2$ & $2$\\
 \hline
 $8$ & $35$ & $35$ & $10$ & $10$\\
 \hline
 $9$ & $70$ & $70$ & $8$ & $8$\\
 \hline
 $10$ & $-126$ & $138$ & $0$ & $4$\\
 \hline
 $11$ & $-252$ & $252$ & $-22$ & $22$\\
 \hline
 $12$ & $462$ & NA & $80$ & $80$\\
 \hline
 $13$ & $924$ & NA & $72$ & NA\\
 \hline
 $14$ & $-1716$ & NA & $0$ & NA\\
 \hline
 $15$ & $-3432$ & NA & $-228$ & NA\\
 \hline
 $16$ & $6435$ & NA & $810$ & NA\\
\hline
\end{tabular}
\caption{The difference between the number of green and red configurations $g-r$ and the number of zero-energy states $N_0$ (determined by exact diagonalization) for different system sizes. Overall, the derived bound for the number of zero modes is found to be very tight in finite systems where it can be independently confirmed by explicit diagonalization (``NA" denotes cases where this was not possible).
\label{tab:gr}}
\end{table}

\FloatBarrier

\section*{Supplementary Notes}

\subsection*{Supplementary Note 1: Hopping coefficients}\label{a:hc}

Constraints are not the only factor that slows down the dynamics and leads to weakly-entangled eigenstates in spectrum. The relative magnitude of the hopping coefficients between different configurations mapped to each other under the action of the Hamiltonian also has important effects. In order to show this, we introduce two additional Hamiltonians
\begin{eqnarray}\label{eq:h1a}
 H_{1\mathrm{a}}&=&-J\sum_{j} \left(b^\dagger_{j}b_{j+1}n^2_{j}+n^2_{j-1}b^\dagger_{j}b_{j-1}\right),\\\label{eq:h1b}
 H_{1\mathrm{b}}&=&-J\sum_{j} \left(b^\dagger_{j}b_{j+1}(1-\delta_{n_{j},0})+(1-\delta_{n_{j-1},0})b^\dagger_{j}b_{j-1}\right).\nonumber\\ \ 
 \end{eqnarray}
 These two Hamiltonians have the same constraints as the Hamiltonian $H_1$ and therefore the same graphs as in Fig.~1 in Section ``Graph structure of the models''. 
 However, the particle number operators $n_{j}$ are squared in $H_{1\mathrm{a}}$ and replaced with delta functions $(1-\delta_{n_{j},0})$ in $H_{1\mathrm{b}}$. This makes the minimal and extended clusters 
 even less connected to the rest of the configurations in the case of $H_{1\mathrm{a}}$ and more connected in the case of $H_{1\mathrm{b}}$.

As anticipated, the revivals become more prominent for $H_{1\mathrm{a}}$, with fidelity peaks reaching more than $95\%$, while the peaks almost disappear for $H_{1\mathrm{b}}$, as illustrated in Supplementary Figure~\ref{fig:dynH1ab}(a). In addition, the entanglement entropy quickly saturates in the case $H_{1\mathrm{b}}$, while the growth is significantly suppressed  in the case of $H_{1\mathrm{a}}$, as can be observed in Supplementary Figure~\ref{fig:dynH1ab}(b). The distribution of entanglement entropy across all eigenstates is also affected by the change of coefficients (not shown). 
$H_{1\mathrm{a}}$ has a spectrum with many low-entropy eigenstates, while the spectrum of $H_{1\mathrm{b}}$ is almost thermal and resembles that of $H_2$.
The probability distribution of consecutive gaps in the energy spectrum of $H_{1\mathrm{a}}$ is close to the Poisson distribution, which
implies that $H_{1\mathrm{a}}$ is almost integrable. On the other hand, the distribution for $H_{1\mathrm{b}}$ is Wigner-Dyson, like in $H_1$.

\subsection*{Supplementary Note 2: Derivation of fidelity in the extended cluster approximation for $\lvert(210)^n\rangle$}\label{a:c2}

For system size $L=3$, the Hilbert space of the extended cluster is spanned by only four configurations: 
\begin{eqnarray}
\begin{pmatrix}
1 \\
0 \\
0 \\
0
\end{pmatrix}
= \lvert300\rangle,\
\begin{pmatrix}
0 \\
1 \\
0 \\
0
\end{pmatrix}
= \lvert210\rangle,\ 
\begin{pmatrix}
0 \\
0 \\
1 \\
0
\end{pmatrix}
= \lvert120\rangle,\ 
\begin{pmatrix}
0 \\
0 \\
0 \\
1
\end{pmatrix}
= \lvert111\rangle.\nonumber\\
\end{eqnarray}
 The Hamiltonian reduced to this subspace is
\begin{eqnarray}
H_{1}^{\tilde{\mathrm{c}}}=-
\begin{pmatrix}
0 & 2\sqrt{3} & 0 & 0\\
2\sqrt{3} & 0 & 2 & 0\\
0 & 2 & 0 & \sqrt{2}\\
0 & 0 & \sqrt{2} & 0
\end{pmatrix}.
\end{eqnarray}
Its eigenvalues are
\begin{eqnarray}
 E_1=-\alpha,\ E_2=\alpha,\ E_3=-\beta,\ E_4=\beta,
\end{eqnarray}
and its eigenvectors
\begin{eqnarray}
\lvert1\rangle=
\begin{pmatrix}
a \\
b \\
c \\
d \\
\end{pmatrix}
,\
\lvert2\rangle=
\begin{pmatrix}
-a \\
b \\
-c \\
d \\
\end{pmatrix}
,\
\lvert3\rangle=
\begin{pmatrix}
-c \\
-d \\
a \\
b \\
\end{pmatrix}
,\
\lvert4\rangle=
\begin{pmatrix}
c \\
-d \\
-a \\
b \\
\end{pmatrix}
,\nonumber\\
\end{eqnarray}
where $\alpha=\sqrt{9+\sqrt{57}}\approx4.06815$, $\beta=\sqrt{9-\sqrt{57}}\approx1.20423$, $a\approx0.591050$, $b\approx0.694113$, $c\approx0.388150$ and $d\approx0.134933$.
There are no simple analytical expressions for the coefficients $a$, $b$, $c$ and $d$.

The configuration $\lvert210\rangle$ 
evolves as
\begin{eqnarray}
\nonumber \lvert\psi^{\tilde{\mathrm{c}}}_1(t)\rangle
\nonumber &=&-2i\left(ab\sin \alpha t+cd\sin \beta t\right)\lvert300\rangle\\
\nonumber &&+2\left(b^2\cos \alpha t+d^2\cos \beta t\right)\lvert210\rangle\\
\nonumber &&-2i\left(bc\sin \alpha t-ad\sin \beta t\right)\lvert120\rangle\\
&&+2bd\left(\ \cos \alpha t-\ \cos \beta t\right)\lvert111\rangle,
\end{eqnarray}
which can also be generalized to larger systems
\begin{eqnarray}\label{eq:psi-c2}
 \lvert\Psi^{\tilde{\mathrm{c}}}_n(t)\rangle&=&\lvert(210)^n(t)\rangle=
2^n\left(b^2\cos \alpha t+d^2\cos \beta t\right)^n\lvert(210)^n\rangle\nonumber\\&&+...
\end{eqnarray}
Finally, the fidelity evolves as
\begin{eqnarray}\label{eq:f-c2a}
 F^{\tilde{\mathrm{c}}}_{n}(t)&=&\lvert\langle\Psi^{\tilde{\mathrm{c}}}_n(0)\lvert\Psi^{\tilde{\mathrm{c}}}_n(t)\rangle\lvert^2\nonumber\\
 &=&4^n\lvert b^2\cos \alpha t+d^2\cos \beta t\lvert^{2n}.
 \end{eqnarray}
The period of revivals is approximately $T\approx\pi/\alpha \approx0.772241$,
and the first peak height exponentially decreases as
\begin{eqnarray}\label{eq:fph}
 F^{\tilde{\mathrm{c}}}_{L=3n}(T)&=&4^n\lvert d^2\cos \frac{\pi \beta}{\alpha }-b^2\lvert^{2n}\approx 0.887017^n\nonumber\\
 &\approx&\mathrm{e}^{-0.119891n}\approx\mathrm{e}^{-0.039964L}.
 \end{eqnarray}
Eqs.~\eqref{eq:psi-c2} and \eqref{eq:f-c2a} are exact for non-translation-invariant initial states, but just an approximation for the translation symmetric case. 
This is due to the fact that different translations of $300$, $210$ and $120$ no longer evolve independently in that case, as they are connected to each other through the configuration $111$.
However, this approximation becomes better with increasing the system size, as the configuration $(111)^n$ becomes further away from the initial state $(210)^n$ and the probability that this configuration will be reached decreases. 

\subsection*{Supplementary Note 3: Bipartite structure of $H_1$}\label{sec:bipartite}

We have shown in Methods 
that the $H_1$ model is bipartite for open boundary conditions irrespective of the system size $L$ parity or for periodic boundary conditions when $L$ is even. 
Here we prove that this property of $H_1$ holds true when $L$ is odd and $\nu=1$.

 Due to the constraints imposed by $H_1$, a particle cannot hop to an empty site to its left.
 At the filling factor $\nu=1$, all configurations except $111...111$ contain at least one empty site. 
 These configurations can be connected to $111...111$ by hoppings only to the right, which is also the shortest possible path $d_a$, defined in Eq.~(17) in Methods. 
 The empty sites can be filled only with particles that come from the site on their left, as hopping from the other side is forbidden. This implies that at least for one pair of adjacent sites (an empty site and the filled one on its right) there will be no particles hopping between them on the path to $111...111$. We can then redefine the numbering of sites to start from the filled site in this pair. 
 This is equivalent to setting the right (filled) site as the first and the left (empty) site as the last site in the chain and imposing open boundary conditions.
 In this way, no two odd sites will be coupled and the argument that the absolute difference between the numbers of particles on even and odd sites can only change by $\pm1$ will still be valid.
 
  Unlike $H_1$, $H_2$ in the same geometry is not bipartite.
 The reason for this is that there are no constraints in the case of $H_2$, so the shortest path to $111...111$ can include hoppings both to the right and to the left, which means that it is not always possible to choose the numbering in such a way that no two odd sites are coupled.
  Because of the open boundary conditions, the Hamiltonian $H_3$ in its largest connected component is also bipartite for all system sizes.

\subsection*{Supplementary Note 4: Zero modes}\label{sec:zero}

An interesting feature of $H_1$ model is the large number of zero energy states in the middle of its spectrum. The number of these zero modes, found by brute force diagonalization, is listed in Supplementary Table~\ref{tab:zm} for different system sizes and momentum sectors. Similar property is found for $H_2$ -- see Supplementary Table~\ref{tab:zm2}, with the notable difference that there are no zero modes when the number of sites $L$ is odd.


The origin of the zero modes is the underlying bipartite structure of the Hamiltonian~\cite{Sutherland86,Inui}. 
As explained in Methods, 
all the basis configurations of the $H_1$ model can be separated into two disjoint classes, and the action of the Hamiltonian $H_1$ only connects configurations in one class to the configurations in the other class, while $H_1$ does not connect configurations within the same class. 
For example, a graph that shows how the configurations for $L=N_\mathrm{p}=4$ are connected is displayed in Supplementary Figure~\ref{fig:L4}.
Here we will refer to the two classes as the ``green'' (even) and the ``red'' (odd) configurations. 
Each basis configuration can be uniquely assigned a green or a red label according to the parity of its distance $d_a$
from the configuration $\lvert111...111\rangle$, Eq.~(17) in Methods. 
If this number is even, the configuration is green, and if it is odd, the configuration is red. 

This separation into two classes is a consequence of the constraints present in the Hamiltonian $H_1$. 
Hamiltonians without such constraints, for example $H_2$ or the standard Bose-Hubbard model, 
do not exhibit this bipartite structure for odd system sizes, see  
Supplementary Note~3 and Methods for more details. In these cases, it is not possible to uniquely determine whether a particular configuration is green or red. 
The lack of bipartite structure is the reason for the absence of zero energy eigenstates of $H_2$ in odd dimensions, which was observed in Supplementary Table~\ref{tab:zm2}. However, the configuration space of $H_2$ is still bipartite in even dimensions, allowing for the existence of some zero modes in those cases.

Low-entropy zero-energy states can be constructed as superpositions of either only green or only red configurations.
For example, in the case of $L=N_\mathrm{p}=4$, the simplest and therefore the lowest-entropy zero mode can be constructed using only two green product states (encircled by a dashed line in Supplementary Figure~\ref{fig:L4})
\begin{eqnarray}
 \lvert \psi_0\rangle=\frac{1}{\sqrt{3}}\lvert1111\rangle-\sqrt{\frac{2}{3}}\lvert2020\rangle_\mathrm{T},
\end{eqnarray}
where $|\ldots\rangle_T$ was defined in Eq.~(24) in Methods. 
There is another zero mode in this case, and it can be formed by adding more green configurations to the superposition.
The number of zero-energy eigenstates is related to the difference between the numbers of green and red configurations~\cite{Sutherland86}, as we will now explain.

As the Hamiltonian $H_1$ only connects green configurations to red configurations and red to green, we can rewrite it in the following way:
\begin{eqnarray}\label{eq:h1rg}
 H_1=\sum_{i,j}c_{ij}\lvert R_i\rangle\langle G_j\lvert+
 \sum_{i,j}c^\dagger_{ij}\lvert G_j\rangle\langle R_i\lvert,
\end{eqnarray}
where $\lvert R_i\rangle$ are the red product states and $\lvert G_j\rangle$ are green. Its square, $H^2_1$, connects green configurations to green and red to red, and it is therefore block diagonal. The blocks are $CC^\dagger$ and $C^\dagger C$, where $C$ is a matrix with the elements $c_{ij}$. The dimensions of $C$ are $r\times g$, where $r$ is the number of red configurations and $g$ of green. $C$ and $C^\dagger$ can be factorized using singular value decomposition. From this structure we can see that the energy spectrum is symmetric around zero and that the minimal number of zero-energy states is $\lvert g-r\lvert$. 
The zero-energy eigenvectors can also be obtained as the ground states of $H^2_1$. 
Similar analysis and counting of the zero modes in PXP model was performed in Ref.~\onlinecite{Turner2017,Schecter2018}.

Supplementary \autoref{tab:gr} shows the difference between the numbers of red and green states $g-r$ for different system sizes and the number of zero-energy states $N_0$ in those systems. 
The number of zero-energy states, found by exact diagonalization,  in all cases satisfies the anticipated inequality $N_0\geq\lvert g-r\lvert$. In fact, the bound is almost always saturated, $N_0=\lvert g-r\lvert$, except when $L=N_\mathrm{p}=4n+2$, $n\in\mathbb{Z}$. 
Interestingly, the minimal number of zero modes $\lvert g-r\lvert$ for $L=N_\mathrm{p}=4n$ is equal to the Hilbert space dimension for $L=N_\mathrm{p}=2n$ (both total and for $k=0$ sector only). 
This leads to the conclusion that the number of zero-energy states grows exponentially with the system size. It can also be noticed that the total difference $g-r$ for $L=N_\mathrm{p}=2n+1$ is always twice the difference for $L=N_\mathrm{p}=2n$.